\documentclass[12pt]{article}

\usepackage{amsmath}
\usepackage{graphicx}
\usepackage{enumerate}
\usepackage{natbib}
\usepackage{url} 

\newcommand{\blind}{0}

\usepackage[T1]{fontenc} % Use 8-bit encoding that has 256 glyphs
\usepackage[utf8]{inputenc} % For Spanish characters
\usepackage[english]{babel} % USEnglish localization
\usepackage[hmargin=2cm,vmargin=2.5cm]{geometry} % Margin sizes
\usepackage{microtype} % Slightly tweak font spacing for aesthetics
\usepackage{fancyhdr} % Allows for nice header and footer
\usepackage{sectsty} % Allows customizing section commands
\usepackage{appendix} % Enables appendices
\usepackage{enumerate} % Custom numerate, useful for i,ii,iii... I,II,III...
\usepackage[colorlinks = true,
            linkcolor = black,
            urlcolor  = blue,
            citecolor = blue,
            filecolor = cyan,
            anchorcolor = blue]{hyperref}
\usepackage[usenames,dvipsnames]{xcolor} % Required for custom colors
\usepackage{lastpage} %  par­tic­u­larly use­ful in the page footer that says: Page N of M.
\usepackage{extramarks} % 
\usepackage{fontawesome} % For web icons

\usepackage{amsthm} % Math Theorems
\usepackage{amssymb} % Math Symbols
\usepackage{amsfonts} % Math Fonts
\usepackage{mathtools} % Extra math tools such as PairedDelimiter
\usepackage{upgreek} % Nice sigma => \upsigma
\usepackage{array} % Enables array features
\usepackage{siunitx} % For SI Unit easy formatting
\usepackage{dsfont} % For math. indicators

\usepackage{float} % Float features
\usepackage{caption,subcaption} % For custom caption environments
\usepackage{wrapfig} % Al­lows fig­ures or ta­bles to have text wrapped around them
\usepackage{tikz} % Diagram and figure creation and rendering
\usepackage{booktabs} % Horizontal rules in tables
\usepackage{float} % Required for tables and figures in the multi
\usepackage{multirow} % Combined rows in tables
\usepackage{multicol} % Combined columns in tables
\usepackage{colortbl} % Color cells
\usepackage{longtable} % Tables than span multipages

\usepackage{listings} % Main package for inserting code
\usepackage[scaled]{beramono} % For using the beramono font
\usepackage[linesnumbered,lined,boxed,commentsnumbered]{algorithm2e} % Allows for algorithm description
\usepackage[noend]{algpseudocode}
\usepackage{datetime} % Date-Time formatting
\usepackage{ulem} % For strikethrough text \st{}
\usepackage[colorinlistoftodos]{todonotes} % useful for leaving todonotes
\usepackage{textcomp} %Text Com­pan­ion fonts

\usepackage{pdfpages} % Insert pdfs
\usepackage{lipsum} % Used for inserting dummy 'Lorem ipsum' text into the template
\usepackage{pdflscape} % Individual horizontal pages
\usepackage[many]{tcolorbox} % Color boxes for comment
\DeclarePairedDelimiter\pa{(}{)}%
\newcommand{\pderiv}[2]{\frac{\partial #1}{\partial #2}}

\newcommand{\Diag}[1]{\mathrm{Diag}\left(#1\right)}
\newcommand{\diag}[1]{\mathrm{diag}\left(#1\right)}
\def\prox{{\mathrm{prox}}}

\DeclareMathOperator*{\sign}{\mathrm{sign}}

\DeclareMathOperator*{\argmin}{\mathrm{argmin}}

\DeclareMathOperator{\R}{\mathbb{R}}

\DeclareMathOperator{\bS}{\mathbb{S}}

\DeclareMathOperator{\NN}{\mathcal{N}}
\def\CC{\mathcal{C}}

\DeclareMathOperator{\LL}{\mathcal{L}}

\DeclareMathOperator{\II}{\mathcal{I}}

\newcommand{\tbeta}{\tilde\beta}
\newcommand{\tgamma}{\tilde\gamma}

\newcommand{\bgam}{\bar{\gamma}}

\newcommand{\half}{\frac{1}{2}}
\newcommand\reals{{{\rm l} \kern -.15em {\rm R} }}

\DeclareMathOperator{\trace}{Tr}

\DeclareMathOperator{\epi}{epi}

\newenvironment{mat}{\left[\begin{array}{ccccccccccccccc}}{\end{array}\right]}
\newcommand\bcm{\begin{mat}}
\newcommand\ecm{\end{mat}}

\newenvironment{rmat}{\left[\begin{array}{rrrrrrrrrrrrr}}{\end{array}\right]}
\newcommand\brm{\begin{rmat}}
\newcommand\erm{\end{rmat}}

\newcommand{\eq}[1]{\begin{equation}\begin{split}#1\end{split}\end{equation}}

\newtheorem{definition}{Definition}
\newtheorem{theorem}{Theorem}

\newtheorem{remark}{Remark}

\def\eR{{\overline\R}}

\def\one{\mathbf{1}}

\def\alf{\alpha}

\def\bgam{{\bar\gamma}}

\def\gam{\gamma}

\def\bgam{{\bar\gam}}
\def\Gam{\Gamma}

\def\del{\delta}

\newcommand\map[3]{#1:#2\rightarrow #3}

\newcommand{\norm}[1]{\left\Vert #1\right\Vert}

\newcommand{\lev}[2]{{\mathrm{lev}_{#1}(#2)}}

\def\emu{{\eta,\mu}}

\begin{document}

\if0\blind
{
  \title{\bf A Relaxation Approach to Feature Selection for Linear Mixed Effects Models}
  \author{Aleksei Sholokhov\thanks{
    Bill and Melinda Gates Foundation}, \hspace{.2cm}\\
    Department of Applied Mathematics, University of Washington,\\
    James V. Burke \\
    Department of Mathematics, University of Washington,\\
    Damian F. Santomauro$^*$, \hspace{.2cm}\\
    School of Public Health, The University of Queensland,\\
    Queensland Centre for Mental Health Research,\\
    Institute of Health Metrics and Evaluation,
    University of Washington, \\
    Peng Zheng$^*$, \hspace{.2cm}\\
    Department of Health Metrics Sciences \&\\
    Institute of Health Metrics and Evaluation,
    University of Washington,\\
    and \\
    Aleksandr Aravkin$^*$, \hspace{.2cm}\\
     Department of Applied Mathematics \& \\
    Institute of Health Metrics and Evaluation, University of Washington \\
    }
  \maketitle
} \fi

\if1\blind
{
  \bigskip
  \bigskip
  \bigskip
  \begin{center}
    {\LARGE\bf Title}
\end{center}
  \medskip
} \fi

\begin{abstract}
Linear Mixed-Effects (LME) models are a fundamental tool for modeling correlated data, 
including cohort studies, longitudinal data analysis, and meta-analysis.  Design and analysis of variable selection methods for LMEs is more difficult than for linear regression because LME models are nonlinear. In this work we 
propose a relaxation strategy and optimization methods that enable a wide range of variable selection methods for LMEs using both convex and nonconvex regularizers, including  $\ell_1$, Adaptive-$\ell_1$, CAD, and $\ell_0$. 
The computational framework only requires the proximal operator for each regularizer to be available, and the implementation is available in an open source \texttt{python} package \texttt{pysr3}, consistent 
with the \texttt{sklearn} standard. 
The numerical results on simulated data sets indicate that the proposed strategy improves on the state of the art for both accuracy and compute time. 
The variable selection techniques are also validated on a real example using a data set on bullying victimization.
\end{abstract}

\noindent%
{\it Keywords:}  Mixed effects models, feature selection, nonconvex optimization 
%\vfill
%
%\newpage
%\spacingset{1.5} % DON'T change the spacing!

\section{Introduction}
Linear mixed-effects (LME) models use covariates 
%(also known as predictors or features) 
to explain the variability of target variables in a grouped data setting. For each group, the relationship between covariates and observations is modeled 
using group-specific coefficients that are linked by a common prior distribution
across all groups, allowing LMEs to borrow strength across groups
in order to estimate statistics for the common prior. 
LMEs are used in settings with insufficient data to resolve each group independently, making them 
fundamental tools for regression analysis in 
population health sciences (\cite{covid2020modeling,murray2020global}), meta-analysis (\cite{dersimonian1986meta, zheng2021trimmed}), life sciences, and 
as well as in many others domains (\cite{zuur2009mixed}). 

Variable selection is a fundamental problem in all regression settings. In linear regression, the LASSO method~\citep{tibshirani1996regression} and related extensions have been widely used. % for this purpose. 
However, variable selection for LMEs is complicated by the nonlinear structure and relative sparsity of the within-group data. 
While standard methods and software are available for linear regression (see e.g. \texttt{glmnet}~\cite{glmnet}), there are few open source libraries for variable selection 
for LMEs. 
Many covariates selection algorithms for LMEs have been proposed over the last 20 years (see the survey \cite{Buscemi2019Survey}), but comparison of 
these strategies and practical application remains difficult. %practical application of these strategies is difficult. 
%Most solutions share the goal of improving selection quality but vary significantly in implementation details. 
Approaches vary by choice of likelihood (e.g. marginal, restricted, or h- likelihood),  
regularizer~(e.g. $\ell_1$~\citep{Krishna2008} or SCAD~\cite{ibrahim2011fixed}), and information criteria \citep{Vaida2005,Ibrahim2011}. 
%and 
%sometimes  estimate effects in stages \citep{Krishna2008}. 
Implementations vary as well, typically using regularizer-specific local quadratic approximations to apply
solution methods for smooth problems (Newton-Raphson, EM, sequential least squares) to fit the original nonsmooth model. 
All of these decisions make it  difficult to compare and evaluate performance of available 
variable selection strategies and to determine which method is best suited for a given task. 
This challenge  is exacerbated by the absence of standardized datasets %for different experimental designs
and open source libraries for each method. 
Our main practical goal to fill this gap by developing a unified methodological framework that
accommodates a wide variety of variable selection strategies based on a set
of easily implementable regularizers, and implemented in an open source library that makes 
it easy to use and to compare different methods. 

\begin{figure}[H]
    \centering
    \includegraphics[width=\textwidth]{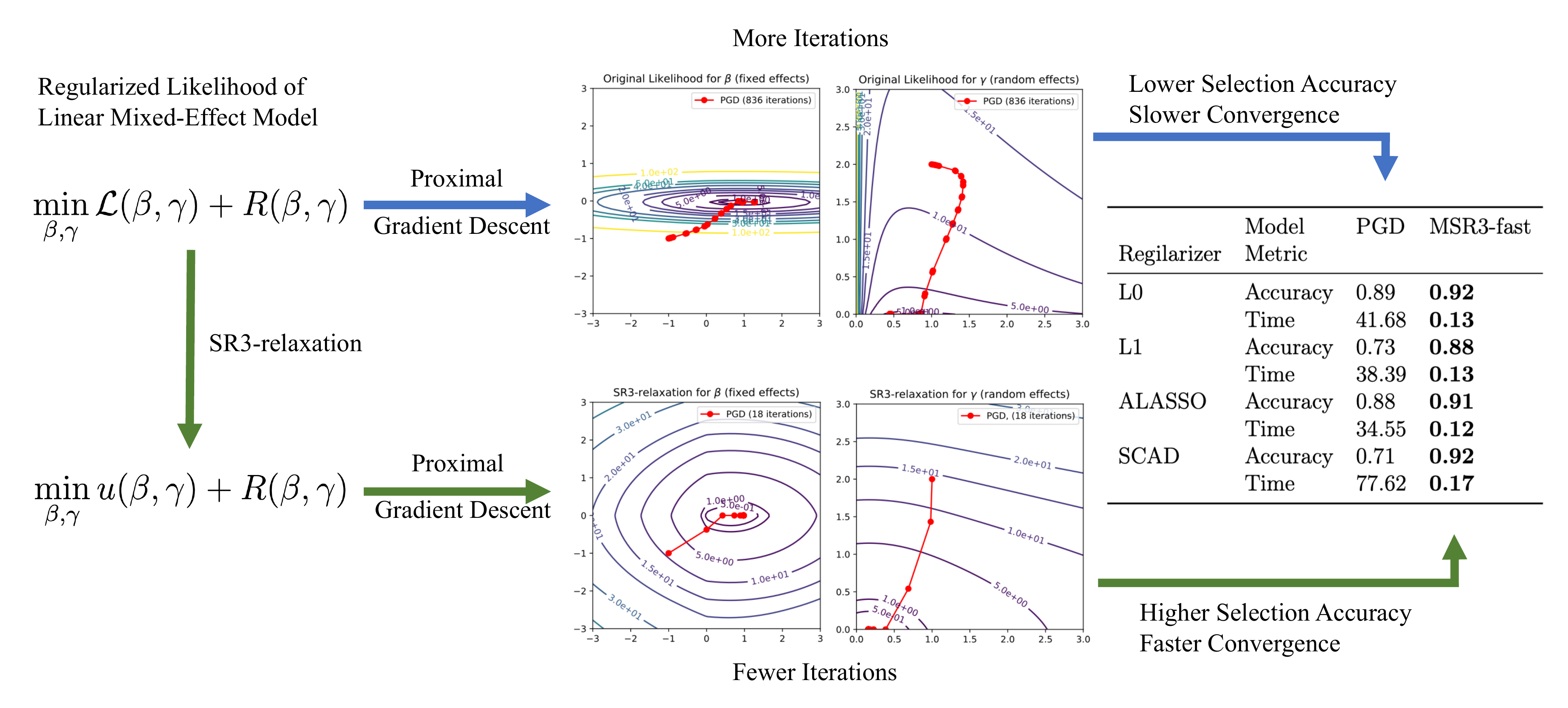}
    \caption{Selection of fixed and random effects for LME likelihoods $\mathcal{L}$ using 
    `regularization-agnostic' framework and its SR3 extension using four regularizers. 
    %standard and SR3-relaxed linear mixed-effect likelihoods with popular sparsity-promoting regularizers for the task of simultaneous 
    %The original likelihood $\LL(\beta, \gamma)$ (Eq. \ref{eq:lmm_objective}) is relaxed with SR3 approach to obtain $u(\beta, \gamma)$ (Eq. \ref{eq:value_function_definition}). Both setups are then solved using the same Proximal Gradient Descent (PGD) algorithm. 
    SR3 relaxation accelerates algorithmic converge  (middle panel), and gives better robustness and improved performance on synthetic problems across regularizers (right panel)}.
    \label{fig:summary}
\end{figure}

In this work we develop a regularization-agnostic covariate selection strategy that (1) is fast and simple to implement, (2) provides robust models, and (3) is flexible enough to support most regularizers currently used in variable selection across different domains.  The baseline approach uses the proximal gradient descent (PGD) method, which has been studied by the optimization community for over 40 years, but has not been widely used in LME covariate selection. We provide proximal operators for commonly used regularizers and show how to apply the PGD method to the nonconvex LME setting. In particular we apply the PGD method to four regularizers, 
including the $\ell_0$ regularizer, which does not admit local quadratic approximations and has not been used before for LME variable selection strategies. 

We also develop a new meta-approach that can improve the performance of LME selection methods for any regularizer. Specifically, we extend sparse relaxed regularized regression (SR3) framework~{(\cite{Zheng2019SR3})} to the LME setting. In linear regression, SR3 accelerates and improves the performance of  regularization strategies by introducing auxiliary variables that decouple the accuracy and sparsity requirements for the model coefficients.
We develop a conceptual and algorithmic approach necessary to extend the SR3 concept to LME.  This development is necessary because the LME problem is nonlinear, nonconvex, and includes constraints on variance parameters.  We show that the new approach yields superior results in terms of specificity and sensitivity of feature selection, and is also computationally efficient.  

All new methods are implemented in an open-source library called \texttt{pysr3}, which fills a gap for python mixed-models selection tools in \texttt{Python} (\cite{Buscemi2019Survey}, Table 3). Our algorithms are  1-2 orders of magnitude faster than available LASSO-based libraries for  mixed effects selection in \texttt{R},  see Table \ref{table:glmmlasso}. \texttt{pysr3} enables a standardized comparison of different methods in the LME setting, and makes both the PGD framework and its SR3 extension available to practitioners working with LME models. 
%\subsection{Notation}
%\begin{enumerate}
%\item sets like $\N$, $\B$, $\bS$, $\SS$.
%\item nonsmooth stuff, subdifferential and normal cones
%\end{enumerate}

\section{Linear Mixed-Effects Models: Notation and Fundamentals}
Mixed-effect models describe the relationship between an outcome variable and its predictors when the observations are grouped, for example in studies or clusters.  To set the notation, consider $m$ groups of observations indexed by $i$, with sizes $n_i$, and the total number of observations equal to $n = n_1 + n_2 + \dots + n_m$. For each group, we have design matrices for fixed features $X_i \in \R^{n_i \times p}$,  and matrices of random features $Z_i \in \R^{n_i \times q}$, along with vectors of outcomes $Y_i \in \R^{n_i}$. 
%Typically, columns of $Z_i$ are a subset of columns $X_i$ but it does not have to be the case in general. 
Let 
{$X = [X_1^T, X_2^T, \dots, X_m^T]^T$ and $Z = [Z_1^T, Z_2^T, \dots, Z_m^T]^T$.} 
Following~\cite{Patterson1971, Pinheiro2000}, we define a Linear Mixed-Effects (LME) model as
\eq{
	\label{eq:lme_setup}
	Y_i & = X_i\beta + Z_iu_i + \varepsilon_i, \quad i = 1 \dots m \\
	u_i & \sim \NN(0, \Gamma),\quad \Gamma \in \bS_{+}^{q} \\
	\varepsilon_i & \sim \NN(0, \Lambda_i), \quad \Lambda_i \in \bS_{++}^{n_i}
}
 where $\beta \in \R^p$ is a vector of fixed (mean) covariates, 
 $u_i \in \R^{q}$ are unobservable random effects assumed to be distributed normally with zero mean and the unknown covariance matrix $\Gamma$, and $\bS_{+}^{\nu}$ 
 and $\bS_{++}^{\nu}$ are the sets of
 real symmetric $\nu\times \nu$ positive semi-definite
 and positive definite matrices, respectively. 
Matrices $Z_i$ encode a wide variety of models, including
random intercepts ($Z_i$ are columns of 1's that add $u_i$ to all datapoints from the $i$th study)
and random slopes ($Z_i$ also scale $u_i$ according to the magnitude of a covarite), see e.g.~\cite{pinheiro2006mixed}. 
%In these typical examples, the columns of $Z_i$ are subsets of the columns
%of $X_i$.  
In our study, we assume that the observation error covariance matrices 
$\Lambda_i$ are given and that the random effects covariance matrix 
is an unknown diagonal matrix, i.e., $\Gamma = \Diag{\gamma}, \ \gamma \in \R^s_+$.
% either
% assumed to have a simple parametric form, such as $\Lambda_i=\sig_i I$
% with $\sig_i$ to be determiined, or given given  
 %; we focus on the latter case below. 
 
Defining group-specific error terms $\omega_i = Z_i u_i + \varepsilon_i$, we get a compact formulation that 
recasts~\eqref{eq:lme_setup} as a correlated noise model:
 \eq{
 \label{eq:lmm_correlated_noise_setup}
	Y_i = X_i\beta + \omega_i,\quad \omega_i \sim \NN(0, \Omega_i(\Gamma)), \quad \Omega_i(\Gamma) = Z_i\Gamma Z_i^T + \Lambda_i.
}
For brevity, we refer to $\Omega_i(\Gamma)$ as just $\Omega_i$.
%without the parentheses. In fact, $\Omega_i$ will be the only terms depending on $\Gamma$ in the majority of expressions including the mixed model's likelihood and its derivatives described below.
The reformulation \eqref{eq:lmm_correlated_noise_setup}
yields the following marginalized  negative log-likelihood function of a linear mixed-effects model~\citep{Patterson1971}:
\eq{
	\label{eq:lmm_objective}
	\LL_{ML}(\beta, \Gamma)  :=
%	 \textabove{$\Delta$}{=}
	 \sum_{i = 1}^m \half(y_i - X_i\beta)^T\Omega_i^{-1}(y_i - X_i\beta) + \half\ln{\det{\Omega_i}}.
}
Maximum likelihood estimates for $\beta$ and $\Gam$ are obtained by
solving the optimization problem
\eq{
	\label{eq:ml_lme_optimization_setup}
	\min_{\beta, \Gamma} & \ \LL_{ML}(\beta, \Gamma)  \quad \mbox{s.t.} \quad \ \Gamma \in \bS_{+}^{q}.
}
At this point, we bring in three basic definitions from variational analysis~\cite{rockafellar2009variational}. 

\begin{definition}[Epigraph and level sets]
The epigraph of a function $f:\mathbb{R}^n\rightarrow \mathbb{R} \cup \{\infty\}$ is defined as
\[
\epi f = \{(x,\alpha) : f(x) \leq \alpha\}. 
\]
For a given $\alpha$, the $\alpha$-level set of $f$ is defined as
\[
\text{lev}_\alpha f = \{x: f(x) \leq \alpha\}.
\] 
\end{definition}

\begin{definition}[Lower semicontinuity and level-boundedness]
A function $f:\mathbb{R}^n\rightarrow \mathbb{R} \cup \{\infty\}$ is lower semicontinous (lsc) when $\epi f$ is closed, 
and level-bounded when all level sets $\text{lev}_\alpha f$ are bounded.  
\end{definition}

\begin{definition}[Convexity]
A function $f:\mathbb{R}^n\rightarrow \mathbb{R} \cup \{\infty\}$ is convex when $\epi f$ is a convex set. Equivalently, 
\[
f(\lambda x + (1-\lambda)y) \leq \lambda f(x) + (1-\lambda) f(y) \quad \forall x,y, \lambda \in (0,1).
\]
\end{definition}

\begin{definition}[Weak convexity]
A function $f:\mathbb{R}^n\rightarrow \mathbb{R} \cup \{\infty\}$ is $\lambda$-weakly convex 
$f(\cdot)+\frac{\lambda}{2}\|\cdot\|^2$ is convex. 
\end{definition}

The negative log likelihood~\eqref{eq:ml_lme_optimization_setup}  is nonlinear and nonconvex, and requires an iterative numerical solver.
However, it is convex with respect to $\beta$, and weakly convex with respect to $\gamma$, with a weak convexity constant $\overline \lambda$
computed in~\cite[Section 5.1]{jimtheory} . The expected value of the posterior mode $\beta$ given $\Gamma$ has a closed form representation of the form
%\eq{
%	\label{eq:beta_formula}
\[
	\beta(\Gamma) = \argmin_{\beta}\LL(\beta, \Gamma) = \left(\sum_{i = 1}^m X_i^T\Omega_i^{-1}X_i\right)^{-1}\sum_{i = 1}^mX_i^T\Omega_i^{-1}y_i.
\] %}
By using the simplification $\Gam=\Diag{\gam}$, we obtain the problem
%{In the meta-analysis setting under study, 
%In our study, we assume that $\Gam$ is a diagonal matrix of the form
%$\Gamma = \Diag{\gamma}, \ \gamma \in \R^s$ with the 
%$\Lambda_i \in \R_{++}^{n_i \times n_i}$ fixed and given 
%%the positive definite constraint from (\ref{eq:ml_lme_optimization_setup}) transforms into a box constraint:
%so that our problem takes the form}
\eq{
	\label{eq:lme_diagonal_setup}
	\min_{\beta\in\R^p, \gam\in\R^q_+} & \ \LL(\beta,\gam)
	:=\LL_{ML}(\beta, \Diag{\gamma}) 
%	\\
%	s.t. & \ \gamma \geq 0\ .
	}
%Under the assumption $\Lambda_i \in \R_{++}^{n_i \times n_i}$ in~\eqref{eq:lme_setup}, 
In this setting, when an
entry $\gamma_j$ takes the  value $0$ the corresponding coordinates of all random effects $u_{ij}$ are identically $0$ for all $i$. 

Verification of the existence to solutions to \eqref{eq:lme_diagonal_setup}
and, more generally, \eqref{eq:ml_lme_optimization_setup} follows from the work of \cite{zheng2021trimmed}.
Standalone proofs for the existence of minimizers are developed in~\cite[Theorem 1]{jimtheory},
and extended to the presence of regularizers in~\cite[Theorem 2]{jimtheory}. %, restated below for convenience. 

%%Here we streamline these techniques and extend to the regularized case, 
%%with basic results given below, with proofs in the appendix.  
%
%\begin{theorem}[Existence of a Minimizer]\label{thm:basic existence}
%Let the assumptions in the statement of problem \eqref{eq:ml_lme_optimization_setup} hold. Then optimal solutions to
%\eqref{eq:ml_lme_optimization_setup} exist.
%\end{theorem}
%
%Feature selection methods  include sparsity-inducing penalties or constraints. See~\cite[Theorem 1]{jimtheory}
%for an existence result under weak assumptions on the regularizer. 
%
%
% 
%
%\begin{theorem}\label{thm:basic existence2}
%Let the assumptions in the statement of problem \eqref{eq:lme_diagonal_setup} hold,
%define $\LL(\beta,\gam):=\LL_{ML}(\beta, \Diag{\gamma})$
%and suppose $\map{\hR}{\R^p\times\R^q_+}{\R\cup\{+\infty\}}$ 
%is lsc and level bounded.
%Then $\LL+\hR$ is level bounded and solutions to
%the following optimization problem exist:
%\eq{
%\label{eq:extended loss}
%\min_{\beta\in\R^p, \gamma\in\R^q_+}  \LL(\beta, \gamma) + 
%\hR(\beta, \gamma) .
%}
%\end{theorem}

In this paper we focus the case where $\Gamma$ is diagonal, 
  %(often referred to as \textit{the diagonal setup}) 
  and all $\Lambda_i$ are known (see \eqref{eq:lme_diagonal_setup}), 
following the meta-analysis use-case~\citep{zheng2021trimmed}. 
Our numerical experiments show that we get better selection methods by trading complexity of the problem \eqref{eq:ml_lme_optimization_setup} for 
for the simplicity and robustness of the problem \eqref{eq:lme_diagonal_setup}.

\subsection{Prior Work on Feature Selection for Mixed-Effects Models}
\label{sec:prior_work}

\begin{figure}[H]
    \centering
    \begin{tabular}{ccc}
        \includegraphics[width=0.3\textwidth]{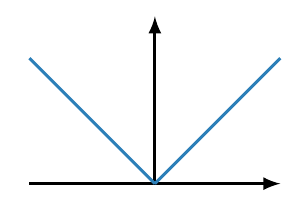}
&
           \includegraphics[width=0.3\textwidth]{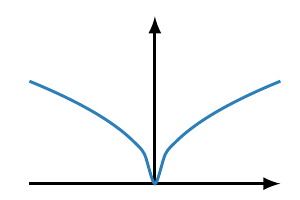}
    &
    \includegraphics[width=0.3\textwidth]{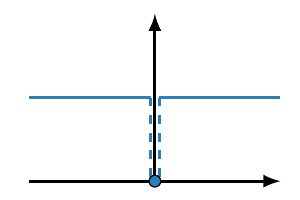}\\
              (a) $\ell_1$  & (b) $\ell_{1/2}$ & (c) $\ell_0$
    \end{tabular}
    \caption{Common convex and non-convex regularizers used for feature selection.}
    \label{fig:regularizers}
\end{figure}

Variable (feature) selection seeks to select or rank the most important predictors in a dataset in order to get a parsimonious model at a minimal cost to prediction quality. 
%The selection process often occurs simultaneously with fitting the model, and a predictor is `selected' if its respective coefficient in the model is non-zero. 
If the desired number of coefficients $k$ is given, then the feature selection problem can be formulated as the minimization of a loss function $f(\theta)$ (e.g. the negative log-likelihood) subject to a zero-norm constraint:
\eq{
	\label{eq:general_feature_selection_zero_norm}
	&\min_{\theta}  f(\theta) \quad \mbox{ s.t. }  \quad \|\theta\|_0 \leq k 
	}
where $\|\theta\|_0$ denotes the number of nonzero entries in $\theta$, see panel (c) of Figure~\ref{fig:regularizers}.

%\begin{figure}
%    \centering
%    \begin{tabular}{cc}
%    \includegraphics[width=0.3\textwidth]{figs/l0_regularizer.pdf}
%    &
%        \includegraphics[width=0.3\textwidth]{figs/l1_regularizer.pdf}\\
%        (a) $\ell_0$ & (b) $\ell_1$ \\
%           \includegraphics[width=0.3\textwidth]{figs/lh_regularizer.pdf}
%    &
%        \includegraphics[width=0.3\textwidth]{figs/scad_regularizer.pdf}\\
%              (a) $\ell_{1/2}$ & (b) Clipped Absolute Deviation (CAD) 
%    \end{tabular}
%    \caption{Common convex and non-convex regularizers used for feature selection.}
%    \label{fig:regularizers}
%\end{figure}

% In this form the problem is equivalent to the Knapsack problem, which is NP-complete. Many techniques based on exhaustive search, also known as subset selection process, were developed, and showed to be effective when the total number of predictors is small (see \cite{Muller2013}). The same setup appears to be intractable in a high-dimension setting due to the exponential growth of the number of subsets to check when $n \to \infty$, however, it's possible to get an approximate solution via relaxation techniques.

The constraint in \eqref{eq:general_feature_selection_zero_norm} is 
combinatorial, and a common workaround is to relax it to a one-norm constraint, 
with $\|\theta\|_1$ equal to the sum of absolute values of the entries of $\theta$. 
The best-known example of this approach is the least absolute square shrinkage operator (LASSO) studied by \cite{Tibshirani1996} for linear regression, 
see panel (a) of Figure~\ref{fig:regularizers}.
% for the least squares model:
% \eq{
% 	\label{eq:lasso_regression}
% 	\min_\theta f(\theta) + \lambda||\theta||_1
 %
 
% LASSO combines the small bias of the least-square estimator and the interpretability of the subset selection method, producing
% %, however, 
% biased estimates for the large coefficients~\citep{Zou2006}.  This approach has been extended to leverage many other regularizers that exhibit useful properties including bias 
%reduction for larger coefficients (SCAD from \cite{Fan2001}), simultaneous selection for highly correlated predictors (Elastic Net from \cite{Zou2005}), or better selection accuracy using non-penalized solution's coordinates as weights (A-LASSO from \cite{Zou2006}).
  
 Feature selection for LMEs is more difficult than for linear regression models. 
 In linear regression the observations are independent, whereas in mixed-effects setup they are generally correlated. 
 In addition, LMEs have both mean effect variables  $\beta$ as well as random variance variables 
 $\Gamma$. % where the groups can have different relative importance. 
% Finally, selecting random features implies that $\Gamma$ will have zero columns and rows thereby forcing the optimization methods to the boundary of the feasible parameter set and making both the theoretical analysis and numerical computations more challenging.  
The shrinkage operator approach for linear regression 
\citep{Tibshirani1996} was first adapted 
to the problem of feature selection for the fixed effects in mixed-effect models by \cite{Lan2006}. 
The removal of a random effect from the model requires the elimination of an entire row and column from $\Gamma$. 
To make the problem more tractable, \cite{Chen2003} reparamtrized $\Gamma$ through a modified Cholesky decomposition $\Gamma(D,L): = DLL^TD$, where $D$ is a diagonal matrix and $L$ is a lower-triangular matrix with ones on the main diagonal, and focused on selecting elements of $D$. 
% In this case, one can select variance parameters by selecting the elements of the diagonal matrix $D$ while treating the non-diagonal entries of $L$ as free parameters. 
Based on this idea, \cite{Krishna2008} extended the Adaptive LASSO regularizer (\cite{Lan2006, Xu2015}) to mixed-effects setting using the objective
   \[
        \LL(\beta, \Gamma(D,L)) + \lambda\left(\sum_{i = 1}^p 
        \left|\frac{\beta_i}{\hat\beta_i}\right| + \sum_{j = 1}^q \frac{D_{ii}}{\hat{D_{ii}}} \right),
   \]
 where $\hat\beta$ and $\hat{D}$ are the solution of a non-penalized maximum likelihood problem. 
 %This strategy has a well known issue: non-penalized estimates $\hat\beta$ and $\hat{D}$ may be inaccessible because the numerical algorithm may fail to converge, especially when the underlying solution is sparse. 
 \cite{Ibrahim2011} use a similar approach, penalizing  non-zero elements $\Gamma_{ij}$ directly. 
 Other methods that use Adaptive LASSO for simultaneous selection 
 of fixed and random effects are \cite{Lin2013, fan2014robust, pan2018simultaneous}. Adaptive LASSO is available to practitioners via \texttt{R} packages \texttt{glmmLasso}\footnote{\href{https://rdrr.io/cran/glmmLasso/man/glmmLasso.html}{https://rdrr.io/cran/glmmLasso/man/glmmLasso.html}} (\cite{groll2014variable}) and \texttt{lmmLasso}\footnote{\href{https://rdrr.io/cran/lmmlasso/}{https://rdrr.io/cran/lmmlasso/}}(\cite{schelldorfer2011estimation}).

 A popular nonconvex regularizer used for feature selection is smoothed clipped absolute deviation (SCAD)~\cite{Fan2001}. 
 %The (non-smoothed) clipped absolute deviation penalty is depicted in panel (d) of Figure~\ref{fig:regularizers}.
% , and the algebraic forms for both CAD and SCAD are given in the appendix. 
% which acts on the first derivative of the penalty function $p_\lambda(\theta)$:
 %  \eq{
  % 		p'_{\lambda}(|\theta|) = \lambda\left\{\ind{\theta \leq \lambda} +  \frac{(a\lambda - \theta)_+}{(a-1)\lambda}\ind{\theta - \lambda} \right\}
  % }
The adaptation of the SCAD penalty to select both fixed and random features in 
linear mixed models was developed by \cite{Fan2012}. SCAD was also used by \cite{chen2015inference} for selecting fixed effects and establishing the existence of random effects in ANOVA-type models. Finally, \cite{Ghosh2018} studied SCAD regularization for selecting mean effects in high-dimensional genomics problems.

To better compare methods, we need to consider the regularization parameter and how it is tuned. 
The output of a shrinkage model depends on this tuning parameter, typically called $\lambda$. The entire range of possible $\lambda$ values is captured by the notion of a ``$\lambda$-path in the model space'', with  the best parameter and the final model chosen using information criteria. According to \cite{Muller2013}, the most widely used information criterion is the marginal AIC criterion (\cite{Vaida2005}):
  \eq{
  \label{eq:vaida_aic}
  AIC = 2\LL(\hat{\theta}) + 2\alpha_n(p+q)
  }
  where $\hat\theta$ includes all the estimated parameters $(\beta, \Gamma)$, and $\alpha_n = n(n-p-q-1)$ for a finite sample case (\cite{Sugiura1978}). 
  %As discussed by \cite{Fang2011}, AIC is asymptotically equivalent to leave-one-out cross-validation, so it can be used for choosing between a finite number of models. AIC, however, is also known to be positively-biased (\cite{Ibrahim2011}). 
  Alternatively, LASSO-type methods (\cite{Krishna2008, Ibrahim2011}) use a BIC-type information criterion:
  \eq{
    BIC = 2\LL(\hat{\theta}) + \log(n)(p+q).
  }
BIC performs well in practice, but does not have theoretical guarantees~(\cite{schelldorfer2011estimation}).
  %One can refer to \cite{Muller2013, Buscemi2019Survey} for a detailed overview of different feature selection approaches.
  
\section{Algorithms for Feature Selection}
\label{sec:pgd_methods}
To develop our  feature selection approach, we add a regularizer to  model \eqref{eq:lme_diagonal_setup}: 
%( $\Lam_i$ are known and given, and $\Gamma$ an unknown diagonal matrix,   
%to be estimated along with the fixed effects $\beta$. 
%and add a regularizer:
%to the objective to obtain the optimization problem described
%in Theorem \ref{thm:basic existence2}:

%Feature selection is imposed by adding
%Consider a mixed-effect model setting described in Eq. (\ref{eq:lme_setup}) with $\Gamma$ being diagonal: $\Gamma = \Diag{\gamma}$.  
%We wish to find a minimizer of the problem 
%\eq{
%	\label{eq:lme_loss_original}
%	\min_{\beta\in\R^p, \gamma\in\R^q_+} & \LL(\beta, \gamma) + R(\beta, \gamma), 
%	}
\eq{%\label{eq:main opt in x over C}
    \label{eq:lme_loss_original_in_x}
    \min_{x } & \LL(x) + R(x)+ \delta_{\CC}(x),
%    \\
%    \text{s.t. } & x \in \CC\ .
}
where $x = (\beta, \gamma)$, 
$\CC:=\R^p\times \R^q_+$,
$\map{R}{\R^P\times\R^q_+}{\eR_+:=\R_+\cup\{+\infty\}}$
is a 
lower semi-continuous (lsc) regularization term, and
$\delta_{\CC}$ is the convex indicator function
\[
    \delta_{\CC}(x) := \begin{cases} 0, &  x \in \CC \\ +\infty, & x \not\in \CC .\end{cases}
\]  
By Theorem \ref{thm:basic existence2}, solutions to \eqref{eq:lme_loss_original_in_x}
always exist.
% or 
% $\LL(\beta, \gamma)=\LL_{REML}(\beta,\Diag(\gamma))$ in \eqref{eq:reml_objective} with 
% \[
% \B=[\zero,\gam_\mmax\one]:=\left\{ \gamma\, |\, 0\le \gamma_i\le \gam_\mmax,\ i=1,\dots,q\right\}
% \]
% for some $\gam_\mmax>0$ where $\one$ is the vector of the appropriate
% dimension each of whose components is one.
% Since the method applies to either likelihood, hereafter we omit the subscripts and use $\LL$ to denote the objective function. %, leaving this choice up to a practitioner.
%For conciseness, define $x = (\beta, \gamma)$ and 
%$\CC:=\R^p\times \R^q_+$.
%such that $\tbeta \in \R^p$ is unconstrained and $\tgamma \in \R^q_+$ is non-negative:
%\eq{
% x\in \CC \textabove{$\Delta$}{=} \{[\beta', \gamma']: \  \beta' \in \R^p, \ \gamma' \in \R^q_+\} \subset \R^{p+q} 
%}
%With this notation, problem \eqref{eq:lme_loss_original} becomes
%\eq{%\label{eq:main opt in x over C}
%    \label{eq:lme_loss_original_in_x}
%    \min_{x \in \CC} & \LL(x) + R(x),
%    \\
%    \text{s.t. } & x \in \CC\ .
%}
%Equivalently, the box constraint can be moved to the loss in a form of an indicator function:
%or equivalently
%\[
%\eq{
%	\label{eq:lme_loss_original}
%    \min_{x} \LL(x) + R(x) + \delta_{\CC}(x)\ ,
%}
%where 
%$x = (\beta, \gamma)$, 
%$\CC:=\R^p\times \R^q_+$,  
%\quad \text{where}\ \, 
%\delta_{\CC}$ is the convex indicator function
%\[
%    \delta_{\CC}(x) := \begin{cases} 0, &  x \in \CC \\ +\infty, & x \not\in \CC .\end{cases}
%\]
The most common regularizers are separable: 
\begin{equation}\label{eq:R sep}
R(x) = \sum_{i = 1}^p r_i(x_i). 
\end{equation}
Typical choices for the component functions $r_i$ are given in Table \ref{table:proxes}.

\subsection{Variable Selection via Proximal Gradient Descent}

Since $\LL$ is differentiable on its domain, 
the Proximal Gradient Descent (PGD) Algorithm \cite{} offers a simple numerical
strategy for estimating first-order stationary points for 
\eqref{eq:lme_loss_original_in_x} when 
the application of the proximal operator
to $\alpha R + \delta_{\CC}$ is 
computationally tractable. 
%
%whenever the application of the proximal operator \cite{}
%to $\alpha R + \delta_{\CC}$ is 
%computationally tractable, a simple numerical method for estimating
%first-order stationary points for \eqref{eq:lme_loss_original_in_x} is
%the Proximal Gradient Descent (PGD) Algorithm. 
The proximal operator for 
$\alpha R + \delta_{\CC}$ is defined as the mapping
\[ %\eq{
    \prox_{\alpha R + \delta_{\CC}}(z) := \argmin_{y\in \CC}\ R(y) + \frac{1}{2\alpha}\|y - z\|_2^2 ,
\] %}
and the PGD iteration is given by
\[
    x^+ = \prox_{\alpha R + \delta_{\CC}}(x - \alpha \nabla \LL(x)),
\]
where $\alpha$ is a stepsize.
When $R(x)$ has the form given in \eqref{eq:R sep}, we have
\[
 \prox_{R}(z) = (\prox_r(z_1), \dots, \prox_r(z_q)) .
\]
\begin{table}[H]
\small
    \centering
    \begin{tabular}{|p{25.4mm}|c|c|}
        \hline
         \!\!Regularizer & $r(x)$, $x \in \R$ & $\prox_{\alpha r}(z)  $ \\
         \hline \hline
         LASSO ($\ell_1$) \newline (\cite{tibshirani1996regression}) & $|x|$ & $\sign(z)(|z| - \alpha)_+$ \\
         \hline
         A-LASSO \newline (\cite{Fan2001}) & $\bar{w}|x|$, $\bar{w} \geq 0$ &  $\sign(z)(|z| - \alpha\bar{w})_+$\\
         \hline
         SCAD \newline (\cite{fan1997comments}) & $\begin{cases} \sigma |x|, & |x| \leq \sigma \\ \frac{-x^2 + 2\rho\sigma x - \sigma^2}{2(\rho - 1)}, & \sigma < |x| < \rho\sigma \\ \frac{\sigma^2(\rho + 1)}{2}, & |x| > \rho\sigma \end{cases}$ & $\begin{cases} \sign(z)(|z| - \sigma\alpha)_+, & |z| \leq \sigma(1+\alpha) \\ \frac{(\rho - 1)z - \sign(z)\rho \sigma\alpha}{\rho - 1 - \alpha}, & \sigma(1+\alpha) < |z| \\
         &\quad\leq \max(\rho, 1+\alpha)\sigma \\ z, & |z| > \max(\rho,1+\alpha)\sigma \end{cases}$ \\
%         \hline
%         $\ell_p$, $0 < p < 1$ & $|x|^p$ & Coordinate Newton (\cite{Zheng2019SR3}) \\
         \hline
         $\delta_{\|x\|_0 \leq k}$  \newline ($\ell_0$ ball) & $\begin{cases} 0, & \#\{|x_i| \neq 0\} \leq k\\ \infty, & \text{ otherwise}\end{cases}$ & keep $k$ largest $|x_i|$, set the rest to 0 \\
         \hline
    \end{tabular}
    \caption{Proximal operators for commonly used sparsity-promoting regularizers. }
    \label{table:proxes}
\end{table}

Table \ref{table:proxes} provides closed form expressions for the proximal operators 
of commonly used regularizers. 
%All regularizers in this list are separable except $R(x) = \delta_{\|x\|_0 \leq k}$, for which the solution also has an analytically closed form.
%Again when $R$ is separable with the $r_i$ coming from Table \ref{table:proxes}, 
%the evaluation of 
For all of these cases,  the following theorem gives closed form expressions for
$\prox_{\alpha R + \delta_{\CC}}(z)$.

\begin{theorem}[$\prox$ for bounded $\gamma$]     \label{thm:prox_of_positive_quadrant}
Consider the regularizers from the Table \ref{table:proxes}. % and let $v = (\beta, \gamma)$. % \in \CC \subset R^{p + q}$.
Add a constraint on $\gamma$ of the form 
\[
0 \leq \gamma \leq \bar\gamma,
\]
which has the positive orthant as a special case (i.e. $\bar\gamma = \infty$). We have the following results. 
%    \begin{enumerate}
%    \item  For $R(x)$ given by LASSO, A-LASSO, CAD, and SCAD, we have  for all $i$ that 
    %\eq{
    %\prox_{\alpha R + \delta_{\CC}}(v) = (\prox_{\alpha r}(v_1), \dots, \prox_{\alpha r}(v_p), \prox_{\alpha r + \delta_{\R_+}}(v_{p+1}), \dots, \prox_{\alpha r + \delta_{\R_+}}(v_{p+q}))
    %}
    %where 
 %   \eq{
 %   \prox_{\alpha r + \delta_{\R_+}}(v_{i}) = \begin{cases} \prox_{\alpha r}(v_i), & v_i \geq 0 \\ 0, & \text{ otherwise} \end{cases} 
%    \item For $R(x) = \delta_{\|x\|_0 \leq k}$, the $\prox_{\alpha R + \delta_{\CC}}(v)$ can be evaluated by taking $k$ largest non-negative coordinates of $v$, and setting the rest to $0$.
 %   \end{enumerate}
%    Next, suppose $\LL = \LL_{REML}$. 
    \begin{enumerate}
    \item  For CAD, SCAD, we have for all $i$ that  
   % \eq{
%    \prox_{\alpha R + \delta_{\CC}}(v) = (\prox_{\alpha r}(v_1), \dots, \prox_{\alpha r}(v_p), \prox_{\alpha r + \delta_{\R_+} + \delta_{\R_{\leq \bar\gamma}}}(v_{p+1}), \dots, \prox_{\alpha r + \delta_{\R_+} + \delta_{\R_{\leq \bar\gamma}}}(v_{p+q}))
 %   }
  %  where 
%    \eq{
\[
    \prox_{(\alpha r + \del_{[0,\bgam]})} %\delta_{\R_+} + \delta_{\R_{\leq \bar\gamma}}}
    (\gamma_{i}) = \begin{cases} 
        \prox_{\alpha r}(\gamma_i), & 0 \leq \gamma_i < \bar\gamma  \\
        \bar\gamma, & \gamma_i \geq \bar\gamma \\
        0, & \text{ otherwise} \end{cases}
\]
%    }
    \item  For LASSO, A-LASSO we have for all $i$ that 
 %   \eq{
  %  \prox_{\alpha R + \delta_{\CC}}(v) = (\prox_{\alpha r}(v_1), \dots, \prox_{\alpha r}(v_p), \prox_{\alpha r + \delta_{\R_+} + \delta_{\R_{\leq \bar\gamma}}}(v_{p+1}), \dots, \prox_{\alpha r + \delta_{\R_+} + \delta_{\R_{\leq \bar\gamma}}}(v_{p+q}))
  %  }
  %  where 
%    \eq{
\[
\prox_{(\alpha r + \del_{[0,\bgam]})}(\gamma_i) 
%    \prox_{\alpha r + \delta_{\R_+} + \delta_{\R_{\leq \bar\gamma}}}(v_{i}) 
    = \begin{cases} 
        \prox_{\alpha r}(\gamma_i), & 0 \leq \gamma_i < \bar\gamma + \alpha \\
        \bar\gamma, & \gamma_i \geq \bar\gamma + \alpha \\
        0, & \text{ otherwise} \end{cases}
\]
%    }
    \item For $R(\cdot) = \delta_{\lev{\norm{\cdot}_0}{k}}$  %\delta_{\|x\|_0 \leq k}$, 
    the $\prox_{\alpha R + \delta_{\CC}}(\gamma)$ can be evaluated by taking $k$ largest coordinates of $\gamma$ such that $0 \leq \gamma_i \leq \bar\gamma$, and setting the 
    remainder to $0$.
    \end{enumerate}
\end{theorem}
The proof of the Theorem \ref{thm:prox_of_positive_quadrant} is provided in Appendix \ref{ch:proxes_appendix}. 
%\textcolor{red}{Aleksei, please check proof of appendix is all in terms of $\gamma$, rather than $v = (\beta, \gamma)$.}
The  PGD algorithm is detailed in Algorithm~\ref{alg:pgd_for_lme}.
{The algorithm's step-size $\alpha$ depends on the Lipschitz constant; an upper-bound has been  estimated as in Appendix \ref{sec:lipschitz_constant}. In practice, $\alpha$ is computed using a line-search, since the available estimate of $L$ is very conservative.}

\smallskip

\begin{algorithm}[H]
\SetAlgoLined
$x = x_0$, $\alpha < \frac{1}{L}$,\text{ where } $\LL$ \text{ is $L$-Lipschitz}\\
 \While{not converged}{
    $x^+ = \prox_{\alpha R + \delta_{\CC}}(x - \alpha \nabla \LL(x))$;\\
 }
 \caption{\label{alg:pgd_for_lme}Proximal Gradient Descent for Linear Mixed-Effect Models}
\end{algorithm}
\medskip

\noindent

The main advantages of Algorithm \ref{alg:pgd_for_lme} are its simplicity and flexibility.
The main loop needs only the gradient and prox operator, and the structure of the algorithm is independent of the choice of regularizer $R$.
Algorithm \ref{alg:pgd_for_lme} converges under weak assumptions, in particular neither the objective nor the regularizer need to be convex~\citep{AB17,attouch2013convergence}. 

%we use it only as 
%a point of reference for the algorithm that is the focus of our study which is presented in
%the Section \ref{sec:sr3_adaptation_to_lme}.
%The interested reader should consult the \textcolor{blue}{references \cite{} for the 
%various possible convergence properties
%of Algorithm \ref{alg:pgd_for_lme}}.

\begin{figure}[h!]
    \centering
    \includegraphics[width=\textwidth]{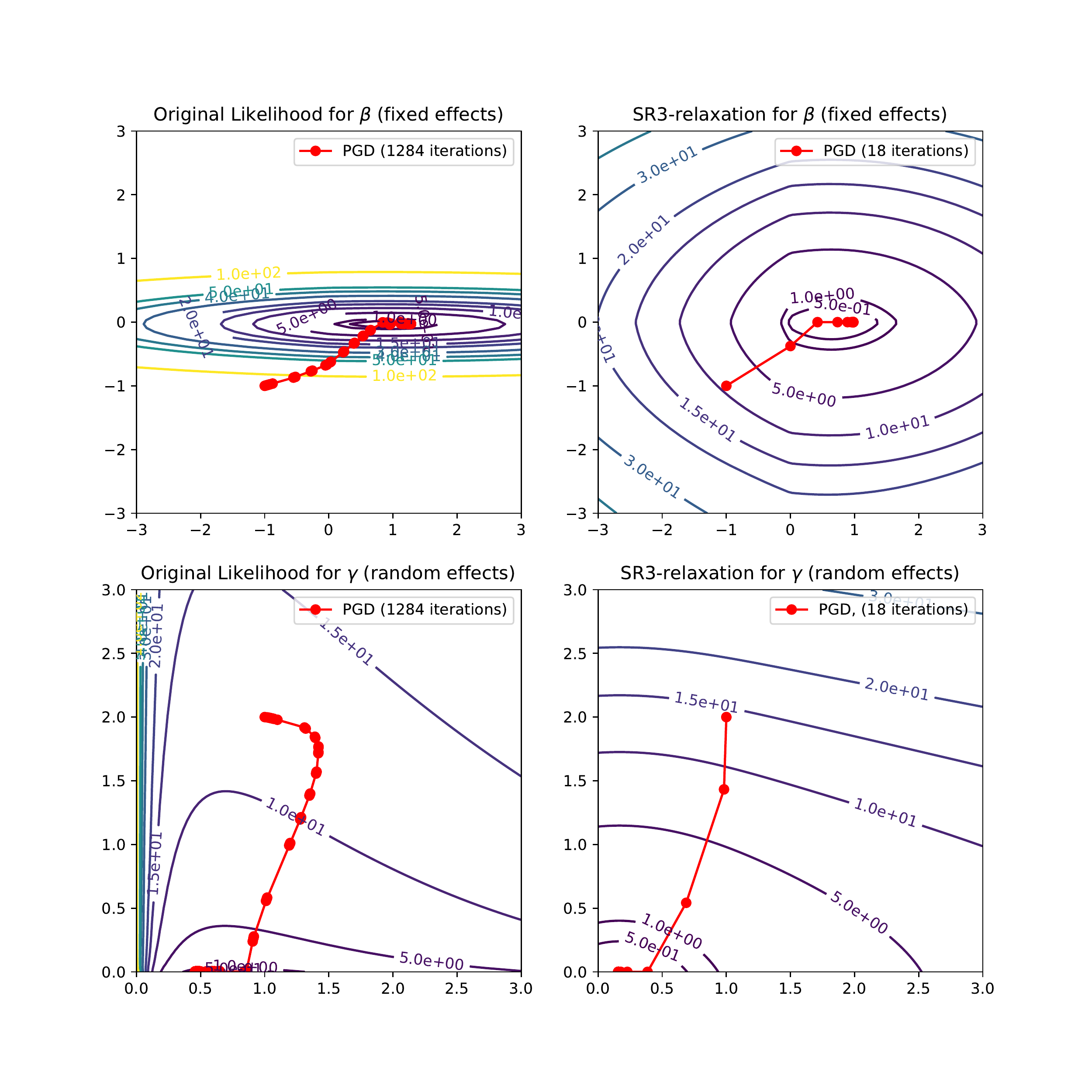}
    \caption{Proximal Gradient Descent (PGD) for~\eqref{eq:value_function_optimization} converges far faster than for~\eqref{eq:lme_loss_original_in_x}, because the SR3-value function $u_\eta$ yields more spherical level-sets than original likelihood for both convex components $\beta$ (first row) and non-convex components $\gamma$ (second row). 
    %The effect is particularly strong for $\gamma$-minimizers near the boundary of the constraint region (bottom-left panel).
    }
    \label{fig:geometric_intuition_sr3}
\end{figure}

\subsection{Variable Selection via MSR3}

To develop an approach that is both more efficient and accurate, we extend the SR3 regularization of~\cite{Zheng2019SR3} to LMEs. 
We call the extension MSR3, since we are focusing on mixed effects models. 
Starting with the regularized likelihood~\eqref{eq:lme_loss_original_in_x} we introduce auxiliary parameters designed to discover the 
fixed and random features: 
\eq{%\label{eq:main opt in x over C}
    \label{eq:lme_loss_sr3}
    \min_{x, w} & \LL(x) + R(w)+ \delta_{\CC}(x) + \kappa_{\eta}(x-w),
%    \\
%    \text{s.t. } & x \in \CC\ .
}
where $\kappa_\eta$ penalizes deviations between $x$ and $w$, and also guarantees that the objective is convex with respect to the $\gamma$
components of $x$:
\begin{equation}
\label{eq:kappa}
\kappa_{\eta} (\beta, \gamma) = \frac{\eta}{2}\|\beta\|^2 + \frac{\overline \lambda + \eta}{2} \|\gamma\|^2.
\end{equation}
Here, $\overline \lambda$ is the weak convexity constant for the objective~\eqref{eq:lme_loss_sr3} with respect to $\gamma$. As $\eta \uparrow \infty$, the extended objective~\eqref{eq:lme_loss_sr3} converges in an epigraphical sense to the original objective~\eqref{eq:lme_loss_original_in_x}. 
However, feature selection accuracy does not require this continuation, and a fixed value, e.g. $\eta=1$, can be used~\citep{Zheng2019SR3}.

To understand the algorithm and logic behind the objective~\eqref{eq:lme_loss_sr3}, we define a value function $u_\eta(w)$ and the solution set $S_\eta(w)$:
\eq{
\label{eq:value_function_definition}
		u_\eta(w) & = \min_x \ \LL(x) + \delta_{\CC}(x) + \kappa_{\eta}(x-w)\\
		S_\eta(w) & = \argmin_x \LL(x) + \delta_{\CC}(x) + \kappa_{\eta}(x-w)
}
Substituting (\ref{eq:value_function_definition}) into (\ref{eq:lme_loss_sr3}) the problem transforms into the problem of optimizing a regularized value function:
\eq{
	\label{eq:value_function_optimization}
	\min_{w} u_\eta(w) + R(w)
}
Thus at a conceptual level, the original regularized likelihood~\eqref{eq:lme_loss_original_in_x} has been transformed through relaxation and partial 
minimization to a mirrored problem~\eqref{eq:value_function_optimization} with the same regularizer. Problem~\eqref{eq:value_function_optimization} 
attempts to find sparse $w$ that are small with respect to $u_\eta$. 

The function $u_\eta$ has a closed form solution for linear regression~\cite{Zheng2019SR3}. However, in both  the linear regression context of~\cite{Zheng2019SR3} 
and in the LME context studied here, we need only compute $S_\eta(w)$ 
in order to optimize~\eqref{eq:value_function_optimization}. The function $u_\gamma$ is well-defined~\cite[Theorem 9]{jimtheory}, differentiable~\cite[Theorem 11]{jimtheory}, and Lipschitz continuous~\cite[Corollary 15]{jimtheory}, and we have a closed form solution for the derivative: 
\[
\nabla u_\eta(w) = \nabla_w k_\eta(x-w)|_{x = S_\eta(w)},   
\]
where $\nabla_w k_\eta(x-w)$ is computed directly from~\eqref{eq:kappa}. 

The advantages of solving (\ref{eq:value_function_optimization}) over (\ref{eq:lme_loss_original_in_x}) come from $u_\eta(w)$ having nearly spherical level-sets while keeping the position of minima close to those of $\LL(x)$. While this effect was extensively studied for a quadratic loss function in the original work (\cite{Zheng2019SR3}), we provide empirical evidence that suggests that it also extends to more general non-linear functions, specifically to the marginal likelihood. In Figure \ref{fig:geometric_intuition_sr3}, we plot the level-sets of $\LL(x) + \|x\|_1$ (left column) and $u(w) + \|w\|_1$ (right column) for the same mixed-effect problem. 
%A more symmetrical 
The more spherical geometry of the latter allows the Algorithm \ref{alg:pgd_for_value_function} to converge in 21 iterations, whereas Algorithm \ref{alg:pgd_for_lme} takes 1284 iterations. The difference is most pronounced when the minimum sits on the boundary of the feasible set, which is always the case for the variable selection problems with true sparse support.

%The value function $u_\eta(w)$ possesses the following properties: 
%
%\begin{theorem}[Properties of value function optimization]
%\label{thm:properties_of_value_function}
%Let $\eta > 0$, $R(x)$ is level-compact. Then
%\begin{enumerate}
%	\item The solutions for (\ref{eq:value_function_optimization}) always exists when $R(w)$ is level-compact \cite[Theorem 6]{jimtheory}.
%	\item The solution set $S_\eta(w)$ is well-defined, single-valued and continuous \cite[Theorem 9]{jimtheory}
%	\item $u_\eta(w)$ is well-defined and continuous \cite[Theorem 9]{jimtheory}.
%	\item $u_\eta(w)$ is differentiable with $\nabla_w u_\eta(w) = \nabla_w \kappa_{\eta}(\bar{x}-w)$ where $\bar{x} = S_\eta(w)$ \cite[Theorem 11]{jimtheory}
%\end{enumerate}
%\end{theorem}

We apply PGD to optimize the value function $u_\eta(w)$ which yields the iteration of the form
\eq{
	\label{eq:pgd_for_value_function}
	w^+ = \prox_{\alpha^{-1}R}(w - \alpha\nabla u_\eta(w))
}
Because of the results stated above, all components of the iteration~(\ref{eq:pgd_for_value_function}) are well-defined. 
To get a deeper intuition for Algorithm~\ref{alg:pgd_for_value} we make the following remark. 
\begin{remark}[Equivalence of Algorithms]
Algorithm~\ref{alg:pgd_for_value} is equivalent to~\eqref{eq:pgd_for_value_function}. 
\end{remark}
\begin{proof}
We extend the relationship studied by~\cite{Zheng2019SR3} to the case of $x = (\beta, \gamma)$. Specifically, we define $\alpha$ to be block-specific: 
	\[
	\alpha = [\underbrace{\eta^{-1}, \dots, \eta^{-1}}_{p}, \underbrace{(\eta + \overline\lambda)^{-1}, \dots (\eta + \overline\lambda)^{-1}}_q]
	\]
so that we have
	\eq{
		\label{eq:grad_value_function_explicit}
		\nabla u_\eta(w) = \alpha \odot (\overline{x} - w)|_{\overline{x} = S_\eta(w)}
	}
where ``$\odot$'' denotes the Hadamard or element-wise product.  Substituting~(\ref{eq:grad_value_function_explicit}) into~(\ref{eq:pgd_for_value_function}), we see that the iteration~(\ref{eq:pgd_for_value_function}) is equivalent to the alternating minimization scheme outlined in the Algorithm~\ref{alg:pgd_for_value}.
\end{proof}
\begin{algorithm}[H]
\SetAlgoLined
$w = w_0$ \\
 \While{not converged}{
    $x^+ = \arg\min_x \LL(x) + \delta_{\CC}(x) + \kappa_{\eta}(x-w)$ \\
    $w^+ = \prox_{\alpha \odot R}(x^+)$
 }
 \caption{\label{alg:pgd_for_value}Proximal Gradient Descent for Value Function}
\end{algorithm}

The alternating strategy is best suited for the simpler context of~\cite{Zheng2019SR3}, where the $x^+$ update has a closed form solution. In the case of~\eqref{eq:lme_loss_sr3}, the regularized loss is convex in $x$ because of the structure of $\kappa$, but is still nonlinear, and requires an iterative algorithm. 

\subsection{Interior Point Method for Inequality Constraints}

In order to solve for the $x^+$ update in line 2 of Algorithm~\ref{alg:pgd_for_value}, we must optimize a convex loss with linear inequality constraints. 
For a fixed $w = (\hat \beta, \hat \gamma)$, we solve 
\begin{equation}
\label{eq:ineqLoss}
\min_{\beta, \gamma} \mathcal{L}(\beta, \gamma) + \kappa_{\eta}(\beta - \hat \beta, \gamma - \hat \gamma) \quad \mbox{s.t.} \quad 0 \leq \gamma. 
\end{equation}
This problem is well suited for an interior point method~\citep{Kojima1991,Nesterov1994,Wright1997}. 
%We now describe how to apply IP methods \teal{to solve the inner problem~\eqref{eq:lme_loss_relaxed} with respect to $x = (\beta, \gamma)$, 
%evaluating the value function $u(w)$ in~\eqref{eq:value_function_definition}}. 
%\eqref{eq:lme_loss_original}. 
% when $\LL=\LL_{ML}$. The case where $\LL=\LL_{REML}$ follows a 
% similar pattern.
%To apply IP methods to our problem class, we first 
First, we relax the inequality constraint $0\le \gamma$ via a log-barrier penalty, obtaining a minimization problem for a new objective $\mathcal{L}_{\mu,\eta}$: 
\eq{
	\label{eq:log_barrier_relaxation_for_ip}
	\min_{\beta, \gamma} &\left\{\mathcal{L}_{\mu,\eta}(\beta, \gamma) :=  \mathcal{L}(\beta, \gamma) + \kappa_{\eta}(\beta - \hat \beta, \gamma - \hat \gamma) - \mu\sum_{i=1}^{q}\ln(\gamma_i) \right\}.
}

The log-barrier penalty approximates the indicator function to the positive orthant as $\mu$ decreases:
\[
\lim_{\mu\downarrow 0} -\mu\ln(\gamma) = \delta_{\mathbb{R}^n_+}(\gamma). 
\]
The penalty (homotopy) parameter $\mu$ is progressively decreased to $0$ as the algorithm proceeds as described below. in the limit minimizing $\mathcal{L}_{\eta,\mu}$ gives the right $x^+$ update in Algorithm~\ref{alg:pgd_for_value}. The existence of solutions for the problem~(\ref{eq:log_barrier_relaxation_for_ip}) for any positive $\mu$ is shown in \cite[Theorem 5]{jimtheory}, and the convergence of solutions to the MSR3 solution as $\mu\downarrow 0$ is shown in~\cite[Theorem 7]{jimtheory}.  
Finally, \cite[Theorem 6]{jimtheory} shows that the MSR3 relaxation is consistent with respect to the barrier, so that as the MSR3 parameter $\eta \uparrow \infty$,
 limit points of global solutions to the former are global solutions to the latter. However, we do not use $\eta$-continuation in the applications considered here.

%The Lagrangian for~\eqref{eq:log_barrier_relaxation_for_ip} is obtained by dualizing the inequality $\gamma \geq 0$ constraint: 
% Define the objective for \eqref{eq:log_barrier_relaxation_for_ip} by
% \eq{
% 	F_\mu(\beta, \gamma) & = \LL(\beta, \gamma) +  \frac{\lambda_b}{2}\|\beta - \tbeta^{(k)}\|^2_2 + \frac{\lambda_g}{2}\|\gamma - \tgamma^{(k)}\|_2^2 - \mu\sum_{i=1}^{q}\log(\gamma_i) %+ v^T\gamma \\
% }

For $\gamma>0$, the necessary optimality conditions 
for $\mathcal{L}_{\mu,\eta}$ in $\gamma$ give us the relation 
%between dual variables $v$ and the primal variables $\gamma$: 
\eq{
	\nabla_\gamma \mathcal{L}_{\mu,\eta}(\beta,\gamma) = 
	%\begin{bmatrix}
\nabla_\gam \LL(\beta,\gam)+(\overline \lambda + \eta) (\gamma - \hat\gamma)
-\mu \Diag{\gam}^{-1}\one
=0,
%		v_1 - \mu/\gamma_1 \\
%		\dots \\
%		v_q - \mu/\gamma_q 
%	\end{bmatrix} = \begin{bmatrix}
%		0 \\
%		\dots \\
%		0 
%	\end{bmatrix} 
%	\implies v_i\gamma_i = \mu \text{ for } i=1,\dots,q.
}
where $\one$ is the vector of all ones of the 
appropriate dimension.
By setting 
\[
v=\nabla_\gam \mathcal{L}_{\mu,\eta}(\beta,\gam)+(\overline \lambda + \eta)(\gamma - \hat \gamma),
\] 
we can rewrite this equation as
%This equation can be written in a matrix form as 
\eq{
	\label{eq:complementary_slackness_kkt}
	v\odot\gamma - \mu\one = 0,
}
where $\one$ is the vector of all ones of the appropriate dimension.
%where ``$\odot$'' denotes the Hadamard (or simply element-wise) product.
%Together with the remaining optimality conditions, we obtain a set of nonlinear equations  that form the the KKT system for~\eqref{eq:log_barrier_relaxation_for_ip}:
The complete set of optimality conditions for 
\eqref{eq:log_barrier_relaxation_for_ip} can now be written as
\eq{\label{eq:IP equations}
	G_{\mu,\eta}(v, \beta, \gamma) & 
%	= \begin{bmatrix}
%		\nabla_v F_\mu \\
%		\nabla_\beta F_\mu \\
%		\nabla_\gamma F_\mu 
%	\end{bmatrix} 
:= \begin{bmatrix}
v\odot \gamma - \mu\one \\
\nabla_\beta \LL(\beta, \gamma) + \eta(\beta - \hat \beta) \\
\nabla_\gamma \LL(\beta, \gamma) + (\overline \lambda + \eta)(\gamma - \hat \gamma) - v
\end{bmatrix} = 0.
}
We then apply Newton's method to~\eqref{eq:IP equations}, 
so in each iteration the search direction $[\Delta v, \Delta \beta, \Delta \gamma]$ solves the linear system
\eq{
	\label{eq:ip_iteration_rs_form}
	\nabla G_{\mu,\eta}(v, \beta, \gamma)\begin{bmatrix}
		\Delta v \\
		\Delta \beta \\
		\Delta \gamma
	\end{bmatrix} = -G_{\mu,\eta}(v, \beta, \gamma).
}
where 
\eq{
	\nabla G_{\mu, \eta}(v, \beta, \gamma) = \begin{bmatrix}
 		\Diag{\gamma} & 0 & \Diag{v} \\
 		0 & \nabla^2_{\beta\beta}\LL + \eta I & \nabla^2_{\beta\gamma} \LL\\
 		-I & \nabla^2_{\gamma\beta}\LL & \nabla^2_{\gamma\gamma}\LL + (\eta + \overline{\lambda})	 I
 	\end{bmatrix}
}
and we have used the fact that  $v\odot \gamma=\Diag{v}\gam=\Diag{\gam}v$. 
The exact formulae for the derivatives of $\LL$ are provided in the Appendix~\ref{appendix:derivatives_of_lmm}.

The general structure of the algorithm is as follows.
Given a search direction
$[\Delta v^{(k)}, \Delta \beta^{(k)}, \Delta \gamma^{(k)}]$, 
choose a step of size $\alpha_k>0$
%\eq{
%    \label{eq:ip_step}
%	v^{(k+1)} & = v^{(k)} + \alpha_k \Delta v \\
%	\gamma^{(k+1)} & = \gamma^{(k)} + \alpha_k \Delta\gamma \\
%	\beta^{(k+1)} & = \beta^{(k)} + \alpha_k \Delta \beta
%}
%with $\alpha_k$ 
so that the update
\[
\begin{pmatrix}
v^{(k+1)}\\ \beta^{(k+1)}\\ \gamma^{(k+1)}
\end{pmatrix}
=
\begin{pmatrix}
v^{(k)}\\ \beta^{(k)}\\ \gamma^{(k)}
\end{pmatrix}
+\alf_k
\begin{pmatrix}
\Delta v^{(k)}\\ \Delta\beta^{(k)}\\ \Delta\gamma^{(k)}
\end{pmatrix}
\]
satisfies the conditions
% chosen to satisfy positivity and sufficient descent conditions:
\eq{
    \label{eq:ip_steplen_conditions}
	\text{\textit{Positivity:}} \qquad & \gamma^{(k+1)} > 0,\ v^{(k+1)} > 0 \\
	\text{\textit{Sufficient Descent:}}\qquad & \| G_\mu(v^{(k+1)}, \beta^{(k+1)}, \gamma^{(k+1)})\| \leq 0.99\|G_\mu(v^{(k)}, \beta^{(k)}, \gamma^{(k)})\| .
}
At each iteration the relaxation parameter $\mu$ is updated by the formula 
\eq{
    \label{eq:ip_mu_update}
    \mu^{(k+1)} = {v^{(k)}}^T\gamma^{(k)}/q,
}
where ${v^{(k)}}^T\gamma^{(k)}$ is the duality gap at
iteration $k$. The algorithm terminates when the criteria 
\eq{
\label{eq:ip_convergence_criterion}
\begin{aligned}
\|G_{\mu,\eta}(v^{(k+1)}, \beta^{(k+1)}, \gamma^{(k+1)})\| &\leq \text{\texttt{tol}}\\
\mu &\leq  \text{\texttt{tol}} 
\end{aligned}
}
are both satisfied, so the interior point problem is nearly stationary, and closely approximates the original problem~\eqref{eq:ineqLoss}.

\paragraph{Positive Approximation of the Hessian}
For many datasets the weak convexity constant $\overline \lambda$ for $\gamma$ can be extremely large or even uncomputable. In this case, setting a smaller value $\overline \lambda$ is likely to force $\nabla^2_{\gamma\gamma}\LL(\beta, \gamma)$ to be negative-definite. Negative definite Hessians may hamper the convergence of second-order methods (\cite{nocedal2006numerical}). We prevent this effect by using the Fisher information matrix regarding to $\gamma$ as the positive definite approximation of the Hessian.
Here, we provide the derivation of the Fisher information matrix formula.From the fact that,
\[
\mathbb{E}\left[(X_i\beta - Y_i)(X_i\beta - Y_i)^T\right] =
\mathbb{E}\left[(Z_i u_i + \epsilon_i)(Z_i u_i + \epsilon_i)^T\right] =
\Omega_i
\]
we know,
\[\begin{aligned}
\mathbb{E}\left[\nabla^2_{\gamma\gamma}\LL(\beta, \gamma)\right]
&= \mathbb{E}\left[\sum_{i = 1}^m \left(Z_i^T\Omega_i^{-1}
(X_i\beta-Y_i)(X_i\beta-Y_i)^T
	\Omega_i^{-1}Z_i\right)
	\circ(Z_i^T\Omega_i^{-1}Z_i)
	-\half(Z_i^T\Omega_i^{-1}Z_i)^{\circ 2}\right]\\
&= \sum_{i = 1}^m \left(Z_i^T\Omega_i^{-1}
\mathbb{E}\left[(X_i\beta-Y_i)(X_i\beta-Y_i)^T\right]
	\Omega_i^{-1}Z_i\right)
	\circ(Z_i^T\Omega_i^{-1}Z_i)
	-\half(Z_i^T\Omega_i^{-1}Z_i)^{\circ 2}\\
&= \sum_{i = 1}^m \left(Z_i^T
	\Omega_i^{-1}Z_i\right)
	\circ(Z_i^T\Omega_i^{-1}Z_i)
	-\half(Z_i^T\Omega_i^{-1}Z_i)^{\circ 2}
= \half\sum_{i=1}^m(Z_i^T\Omega_i^{-1}Z_i)^{\circ 2}
\end{aligned}\]
And we use $\mathbb{E}\left[\nabla^2_{\gamma\gamma}\LL(\beta, \gamma)\right]$
as the positive definite approximation of
$\nabla^2_{\gamma\gamma}\LL(\beta, \gamma)$.

\subsection{Relaxation and Efficient Algorithms: MSR3 and MSR3-Fast }

While algorithm~\eqref{alg:pgd_for_value} is modular, it requires solving a nonlinear optimization problem in $x = (\beta, \gamma)$ for each single update
of $w = (\hat \beta, \hat \gamma)$. To make the implementation as efficient as possible, we designed a more balanced updating scheme, that 
alternates Newton iterations as described in the interior point algorithm with $w$ updates. We update $w$ whenever we are sufficiently close 
to the `central path' in the interior point method, a condition that can be checked rigorously using optimality conditions.

\label{appendix:pseudocode}
\begin{algorithm}[H]
\SetAlgoLined
$\alpha \leftarrow 1$, $\texttt{progress}\leftarrow \textbf{True}$ \\
$\beta^+, \tbeta^+\leftarrow\beta_0$; 
\quad $\gamma^+, \tgamma^+\leftarrow\gamma_0$;  
\quad $v^+ \leftarrow 1 \in \R^q$; 
\quad  $\mu \leftarrow \frac{{v^+}^T\gamma^+}{10 q}$\\
 \While{\texttt{iter} $<$ \texttt{max\_iter}  \ and \ $\|G_\emu(\beta^+, \gamma^+, v^+)\|$ $>$ \texttt{tol}   \ and  \ \texttt{progress} \\}{
    $\beta \leftarrow \beta^+$; \quad $\gamma \leftarrow \gamma^+$; \quad $\tbeta \leftarrow \tbeta^+$; \quad $\tgamma \leftarrow \tgamma^+$ \\
%    $A \leftarrow \nabla G_\emu((\beta, \gamma, v), (\tbeta, \tgamma))$\\
  %  $b \leftarrow G_\emu((\beta, \gamma, v), (\tbeta, \tgamma))$\\
    $[dv, d\beta, d\gamma] \leftarrow  \nabla G_\emu((\beta, \gamma, v), (\tbeta, \tgamma))^{-1}  G_\emu((\beta, \gamma, v), (\tbeta, \tgamma))$ \tcp*[f]{Newton Iteration}\\ 
    $\alpha \leftarrow 0.99\times\min\left(1, -\frac{\gamma_i}{d\gamma_i}, \forall i :\ d\gamma_i < 0\right)$\\
    $\beta^+ \leftarrow \beta + \alpha d\beta$; \quad $\gamma^+ = \gamma + \alpha d\gamma$; \quad  $v^+ \leftarrow v + \alpha dv$\\
    \If{$\|\gamma^+\odot v^+ - q^{-1}{\gamma^+}^Tv^+ \mathbf{1}\| > 0.5q^{-1}{v^+}^T\gamma^+$}{
  %  	\tcp*[h]{Not in the neighborhood of the central path yet}\\
    	continue \tcp*[f]{Keep doing Newton iterations}\\
    }
    \Else{ 
        $\tbeta^+ = \prox_{\alpha R}(\beta^+)$;
        \    $\tgamma^+ = \prox_{\alpha R + \delta_{\R_+}}(\gamma^+)$; 
        \    $\mu = \frac{1}{10}\frac{{v^+}^T\gamma^+}{q}$ \tcp*[f]{Near central path} 
    }
%	\tcp*[h]{Keep iterating until convergence} \\
    \texttt{progress} = ($\|\beta^+ - \beta\| \geq \text{tol}$ or $\|\gamma^+ - \gamma\|  \geq \text{tol}$ or $\|\tbeta^+ - \tbeta\| \geq \text{tol}$ or $\|\tgamma^+ - \tgamma\| \geq \text{tol}$)
 }
 \Return{$\tbeta$, $\tgamma$}
 \caption{\label{alg:pgd_for_value_function}MSR3 (Proximal Gradient Descent for the Value function)}
\end{algorithm}

%The convergence analysis of Algorithm~\ref{alg:pgd_for_value_function} is developed in~\cite[Lemma 12, Theorem 13]{jimtheory}. 

%\input{04Theory}

\section{Verifications}
\label{sec:applications}

\subsection{MSR3 for Covariate Selection}
\label{ch:sr3_l1}

\begin{table}
\centering
\begin{tabular}{lllll}
\toprule
     & Model &    PGD &    MSR3 & MSR3-fast \\
Regilarizer & Metric &        &        &       \\
\midrule
L0 & Accuracy &   0.89 &   \textbf{0.92} &  \textbf{0.92} \\
     & Time &  41.68 &  88.54 &  \textbf{0.13} \\
L1 & Accuracy &   0.73 &   \textbf{0.88} &  \textbf{0.88} \\
     & Time &  38.39 &   9.13 &  \textbf{0.13} \\
ALASSO & Accuracy &   0.88 &   \textbf{0.92} &  0.91 \\
     & Time &  34.55 &  65.19 &  \textbf{0.12} \\
SCAD & Accuracy &   0.71 &   \textbf{0.93} &  0.92 \\
     & Time &  77.62 &  84.67 &  \textbf{0.17} \\
\bottomrule
\end{tabular}
\caption{\label{table:comparison_of_algorithms} Comparison of performance of algorithms measured as accuracy of selecting the correct covariates and run-time. The L0 strategy stands out 
over other standard regularizers. MSR3 improves performance significantly for all regularizers, while MSR3-fast improves convergence speed while preserving the 
accuracy of MSR3.  
More detailed results are in the Table \ref{table:detailed_comparison_of_algorithms} of Appendix \ref{ch:appendix_detailed_comparison}.}
\end{table}

\begin{figure}
    \centering
	\includegraphics[width=1.0\textwidth]{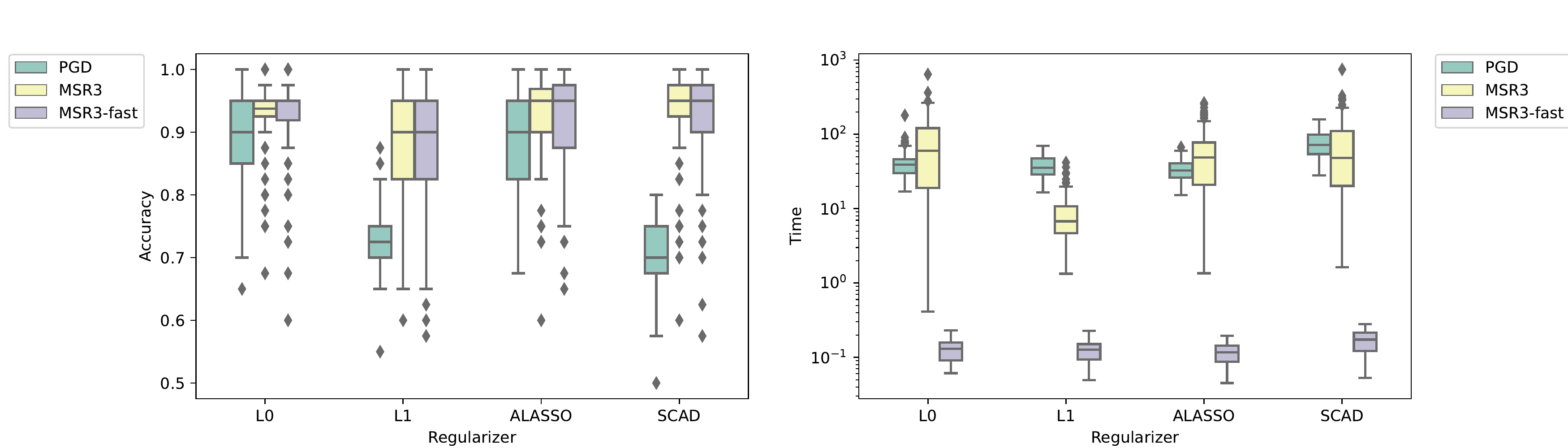}
	\caption{\label{fig:bullying_data_fixed_feature_selection} Feature selection accuracy and execution time in seconds for PGD and MSR3 with various regularizers. MSR3-Fast has the same accuracy as 
	MSR3 and significantly decreases computation time.}
\end{figure}

In this section we show the performance of the `regularization-agnostic' feature selection method, and also show that MSR3 improves performance across different regularization strategies. Specifically, we compare the feature selection accuracy and numerical efficiency of Algorithm  \ref{alg:pgd_for_lme} and \ref{alg:pgd_for_value_function} with LASSO, A-LASSO, CAD, SCAD, and L0. 

%\paragraph{LASSO path} The LASSO path is a set of solutions $x$ parametrized by $\lambda$, where $\lambda$ is sweeping over its range from $\lambda = 0$ to $\lambda \to \infty$, until no parameters are included in the model. It's known that a classic LASSO tends to include false positives early along the path (\cite{Su2017}).

\paragraph{Problem setup.} In this experiment we take the number of fixed effects $p$ and random effects $q$ to be $20$. We set $\beta = \gamma = [\frac{1}{2}, \frac{2}{2}, \frac{3}{2}, \dots, \frac{10}{2}, 0, 0, 0, \dots, 0]$, i.e. the first 10 covariates are increasingly important and the last 10 covariates are not. The data is generated as 
\[
\begin{aligned}
y_i &= X_i\beta + Z_iu_i + \varepsilon_i, \quad  \varepsilon_i \sim \NN(0, 0.3^2 I) \\
X_i &\sim \NN(0, I)^p, \quad Z_i = X_i \\ 
u_i& \sim \NN(0, \Diag{\gamma})\\ 
\end{aligned}
\]

\begin{wrapfigure}{r}{7.5cm}
		\includegraphics[width=7.5cm]{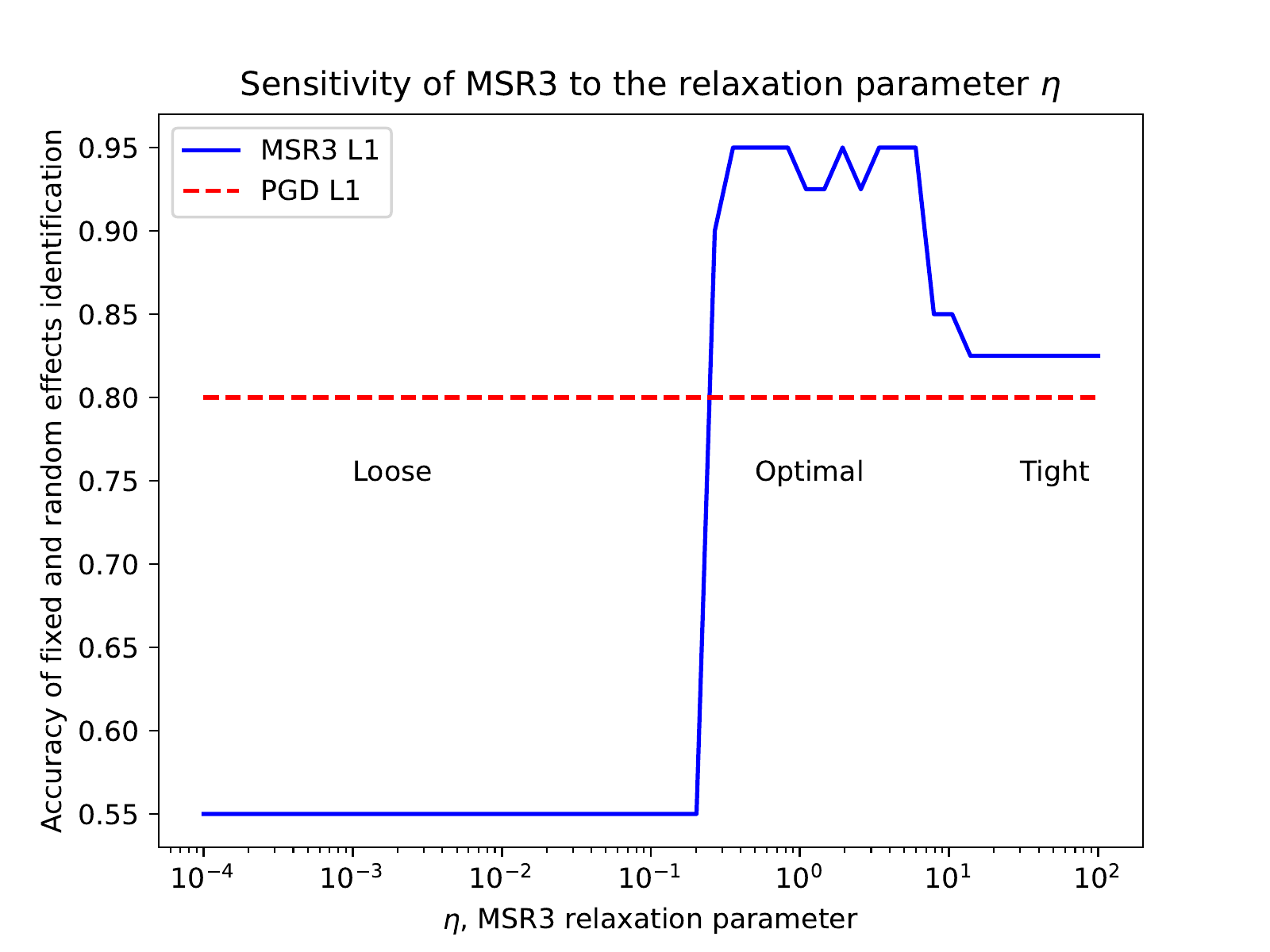}
		\caption{Dependence of model performance on the relaxation $\eta$ for a sample problem.}\label{fig:eta}
	\end{wrapfigure} 
	
The data is generated for 9 groups, with the sizes of $[10, 15, 4, 8, 3, 5, 18, 9, 6]$ to capture a variety of group sizes. To estimate the uncertainty bounds, each experiment is repeated 100 times.
	
\paragraph{Parameter selection.} The regularization coefficient $\lambda$ and the relaxation coefficient $\eta$  were chosen to maximize a classic BIC criterion from \cite{Jones2011}. We used a golden bisection search to select between candidate $\lambda$ parameters and a grid search to select between $\eta$ candidate parameters.

	Figure \ref{fig:eta} shows the dependence of accuracy on the values of $\eta$ for the first problem from our test set. There are three distinct regions, corresponding to loose, optimal and tight levels of relaxations. When $\eta$ is small the relaxation term dominates, and the training does not progress far from the initial point (a fully dense vector \textbf{1} in this case). When the relaxation is tight, the level-sets of the problem converge to those of the original problem, and thus the minimizer. For the values in between, the relaxation significantly improves the model's accuracy. 
These results are consistent with experiments in the sparse linear regression setting~\cite{Zheng2019SR3}. 

\paragraph{Results.}
The results are presented in the Table \ref{table:comparison_of_algorithms}. SR3 improves the selection accuracy of most regularization techniques from Table \ref{table:proxes}, showing a near-perfect performance, while converging two orders of magnitude faster in wall-clock time. 

\paragraph{Comparison to other libraries}
We compare performance of \texttt{pysr3} to the performance \texttt{R} packages \texttt{glmmLasso}\footnote{\href{https://rdrr.io/cran/glmmLasso/man/glmmLasso.html}{https://rdrr.io/cran/glmmLasso/man/glmmLasso.html}} (\cite{groll2014variable}) and \texttt{lmmLasso}\footnote{\href{https://rdrr.io/cran/lmmlasso/}{https://rdrr.io/cran/lmmlasso/}}(\cite{schelldorfer2011estimation}) -- the functionally closest libraries available\footnote{The authors used \cite[Table 3]{Buscemi2019Survey} as a reference list of feature selection libraries. Out of 17 entries mentioned the implementation language, the working libraries were available for \texttt{lmmlasso}, \texttt{glmmlasso}, \texttt{fence}\footnote{\href{https://rdrr.io/cran/fence/}{https://rdrr.io/cran/fence/}} (\cite{jiang2008fence}) and \texttt{PCO} (\cite{lin2013pco}) libraries. \texttt{fence} caused a memory overflow on the experimental system during the performance evaluation on the datasets described above. \texttt{PCO} was not evaluated as it did not support the datasets with the total number of random effects exceeding the number of objects.} online. We evaluate their performance on the same set of problems and the parameters selection procedure as described above and compare it to \texttt{SR3+LASSO}. We tuned the hyperparameters of \texttt{glmmLasso} and \texttt{lmmLasso} by minimizing the BIC scores provided by the libraries. The results are presented in Table \ref{table:glmmlasso}. Overall, SR3 executes, on average, 5 times faster in wall-clock time than \texttt{glmmLasso} and 60 times faster than \texttt{lmmLasso} and shows much higher accuracy of selecting correct fixed and random effects simultaneously. \texttt{lmmLasso} supports the diagonal specification of $\Gamma$ which translates into a competitive quality of selecting random effects. However, while finding sparse fixed effects, \texttt{lmmLasso} provides dense solutions for fixed effects $\beta$. Importantly, the accuracy of \texttt{glmmLasso} is likely skewed downwards due to its BIC selection criterion choosing dense ultimate models and inability to constraint the covariance matrix $\Gamma$ to be diagonal. 

\begin{table}[h]
\centering
\begin{tabular}{lrrrr}
\toprule
Algorithm & Units (perc. / 100 runs) &         MSR3-Fast ($\ell_1$)&         glmmLasso &            lmmLasso \\
\midrule
Accuracy &\% (5\%-95\%)    &        {\bf 88 (72-98)} &        48 (42-55) &          66 (55-73) \\
FE Accuracy &\% (5\%-95\%) &      {\bf 86 (64-100)} &        52 (40-66) &          47 (45-55) \\
RE Accuracy &\% (5\%-95\%) &       {\bf 91 (74-100)} &        45 (45-45) &         84 (55-100) \\
F1          &\% (5\%-95\%)&        {\bf 89 (73-97)} &        63 (60-66) &           65 (0-77) \\
FE F1       &\% (5\%-95\%)&       {\bf 88 (69-100)} &        64 (57-70) &           57 (0-64) \\
RE F1       &\% (5\%-95\%)&       {\bf 90 (73-100)} &        62 (62-62) &          78 (0-100) \\
Time        &sec. (5\%-95\%)&  {\bf 0.19 (0.14-0.24)} &  1.37 (0.78-1.89) &  11.51 (5.35-23.66) \\
Iterations  & num. (5\%-95\%)&        {\bf 34 (28-45)} &        50 (33-77) &             - \\
\bottomrule
\end{tabular}
\caption{\label{table:glmmlasso} Comparison of performance of MSR3-Fast for $\ell_1$ regularizer vs \texttt{glmmLasso}. MSR3-Fast executes 5 times faster in wall time and has higher accuracy of selecting correct covariates. Importantly, the accuracy of \texttt{glmmLasso} is likely skewed downwards due to BIC selection criterion choosing dense ultimate models and due to the missing option to constrain the matrix $\Gamma$ to be diagonal. \texttt{lmmLasso} supports the diagonal specification of $\Gamma$ which translates into a competitive quality of selecting random effects. However, while finding sparse fixed effects, \texttt{lmmLasso} provides dense solutions for fixed effects $\beta$.}
\end{table}

\subsection{Experiments on Real Data}

\begin{figure}
    \centering
	\caption{\label{fig:bullying_data_random_feature_selection}Validation of Random Feature Selection for Bullying Data from GBD 2020. 
	Left panel shows coefficient paths across numbers of nonzero covariates allowed in the model using the $\ell_0$ regularizer. 
	Right panel shows evaluation of each choice, evaluated against expert knowledge from past datasets. 
	The algorithm picks seven historically significant covariates and two historically insignificant, for the model selected using the BIC criteria. 
	See the Appendix \ref{appendix:bullying_covariates} for covariates description and assessment of significance.}
	\includegraphics[width=1\textwidth]{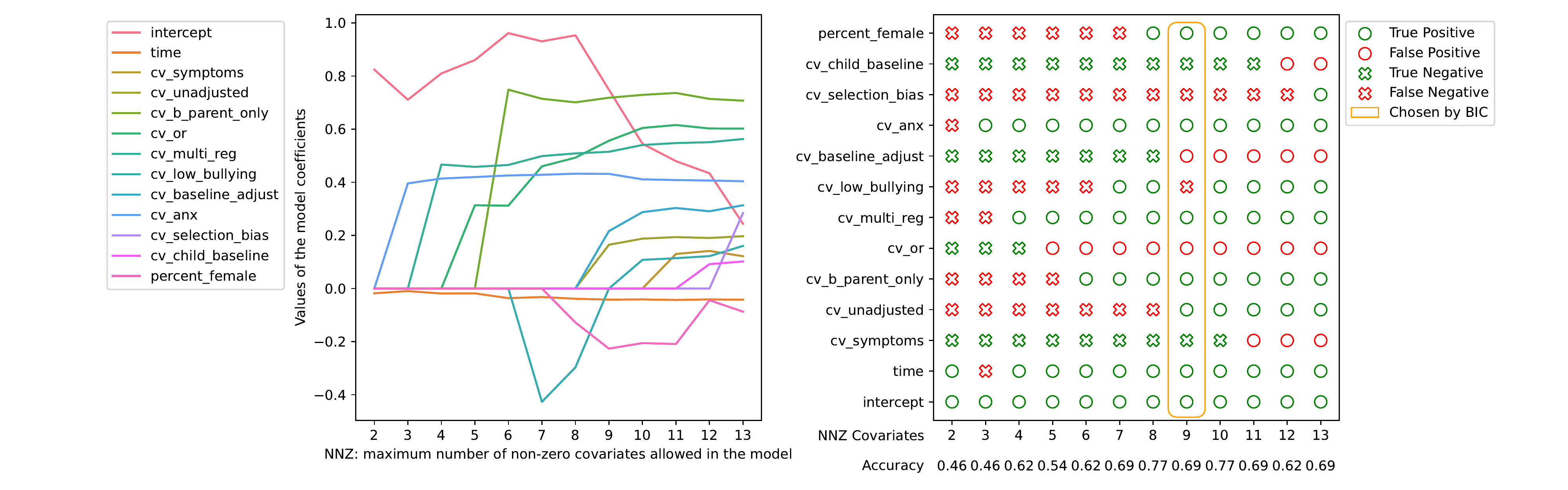}
\end{figure}

In this section we validate the SR3-empowered $\ell_0$-regularized mixed-effect model ($R(x) = \delta_{\|x\|_0 \leq k}$ from Table \ref{table:proxes}) by using it to identify the most important covariates in real data on relative risk of anxiety and depressive disorders depending on the exposure to bullying in young age\footnote{Institute for Health Metrics and Evaluation (IHME). Bullying Victimization Relative Risk Bundle GBD 2020. Seattle, United States of America (USA), 2021.}. This research has been a part of Global Burden of Diseases study for the last several years. The end goal  is to estimate the burden (DALYs) of major depressive disorder (MDD) and anxiety disorders that are caused by bullying. For this risk factor, the exposure is primarily concentrated in childhood and adolescents, but the risk for MDD and anxiety disorders is anticipated to continue well into adulthood. This elevated risk is, however, expected to decrease with time as other risk factors come into play in adulthood (unemployment, relationship issues, etc.). To accommodate this, the research team uses the models which estimate the relative risk (RR) of MDD and anxiety disorders among persons exposed to bullying depending on how many years it has been since the first exposure. Studies informing the model were sourced from a systematic review and consist of longitudinal cohort studies. They measure exposure to bullying at baseline, and then follow up years later and assess them for MDD or anxiety disorders. The detailed description of the covariates can be found in Appendix \ref{appendix:bullying_covariates}.

The feature selection process is illustrated on Figures \ref{fig:bullying_data_fixed_feature_selection} and \ref{fig:bullying_data_random_feature_selection}. 
%Since there is no prior on $k$ -- the number of features to keep in the model -- 
Here, the BIC criterion from \cite{Jones2011} was used to make a choice on $k$ (see the plot on the first row, column two on Figure \ref{fig:bullying_data_fixed_feature_selection}). We see that there is a minimum around $k=4$, although the minimum is shallow so $k=5$ can also be considered. For the $k=4$ case, the selected 
covariates (\texttt{intercept}, \texttt{time}, \texttt{cv\_threshold\_bullying}, \texttt{cv\_b\_parent\_only}) 
are known as important and were used in the analysis in previous years of GBD. For the $k=5$ case, 
%five covariates (in addition to \texttt{intercept} and \texttt{time}) were selected by \ouralgo: \texttt{cv\_unadjusted}, \texttt{cv\_threshold\_bullying}, \texttt{cv\_b\_parent\_only}, \texttt{cv\_anx}, and \texttt{percent\_female}. 
%in addition to the covariates already known to be important, 
the algorithm also selects \texttt{cv\_or} and \texttt{cv\_child\_baseline}, which were not used before. 

\subsection{Software Implementation}
In order to ensure reproducibility of our research, all developed algorithms have been implemented as a part of \texttt{pysr3}\footnote{Available at \href{https://github.com/aksholokhov/pysr3}{https://github.com/aksholokhov/pysr3}} library. This library implements functionality for fitting linear mixed models and selecting covariates. The user interface was designed to be fully compliant with the standards\footnote{\href{https://scikit-learn.org/stable/developers/develop.html}{https://scikit-learn.org/stable/developers/develop.html}} of \texttt{sklearn} library to minimize learning time.

\section{Discussion}
\label{sec:discussion}

    In this paper, we developed and implemented a first-order variable selection framework for LMEs that handles convex and nonconvex regularizers. We also showed that MSR3 relaxation, a regularizer-agnostic transformation of the likelihood, improves the covariates selection accuracy of a wide group of popular sparsity-promoting regularizers. The fact that relaxation  improves accuracy, rather than just serving as a means to solve the original problem,
     is a very interesting  result that can be studied in future work. 
    
    Since the LME relaxation does not have a closed form, we used an interior method to evaluate the requisite value function. We also developed 
    a more efficient version of the algorithm  (MSR3-Fast) that interleaved interior point iterations with updates of the auxiliary variables, and this method was chosen for the 
    open source library \texttt{pysr3} that we provided. 
    %convergence of this method, and illustrate its performance on synthetic and real-world data. 
    %The relaxation shares minima with the original function while having more symmetrical level-sets which accelerate the convergence of gradient-based methods. 
    Numerical experiments on synthetic data showed that the MSR3 approach for variable selection extends regions of hyper-parameter values where the highest accuracy is achieved, making it easier for information criteria to select the best model.  The variable selection library for the accelerated method MSR3-Fast is much faster than currently available software, and allows the MSR3 approach to be easily applied to a range of regularizers that have prox operators available.    

    The main analytic limitations of the proposed method stem from a lack of an analytical representation of the value function in the MSR3 relaxation for LMEs. In contrast to linear regression settings, where the CG method can be efficiently used to evaluate the value function~\cite{Baraldi2019},    the nonlinear optimization problem required for LMEs is more difficult. Using Hessian information makes each iteration computationally efficient but limits the problem size. At the same time, switching to first-order methods for the inner problem inside the relaxation may be prohibitively slow. One way to mitigate the problem would be to build more efficient upper-bound models for the value function that can be evaluated effectively.
    
    The suggested methodology can be expanded to a wider class of models. In particular, one can extend MSR3 to the setting of non-linear mixed-effect models or generalized linear mixed models, which are known to be challenging setups for covariate selection tasks. Both of these problem classes face similar challenges with respect to optimization strategies for highly nonlinear objective functions that arise when we consider marginal likelihoods in these settings. On the other hand, the SR3 strategy would allow any fast solver to be extended to variable selection task through the relaxation reformulation, analogously to what was done here for LMEs.

\bibliographystyle{Chicago}

\bibliography{bibliography}

\clearpage

\appendix

\section{Additional derivations}
\subsection{Derivatives of Marginalized Log-likelihood for Linear Mixed Models}
\label{appendix:derivatives_of_lmm}
For conciseness, let us define the mismatch $\xi_i = Y_i - X_i\beta$. %We also omit the dependence on $\beta$, as it's fixed at this point. 
The loss function \ref{eq:lmm_objective} takes the form 
\eq{
	\label{eq:lmm_simplified_gamma}
	\LL\pa{\gamma} = \sum_{i = 1}^m\half\xi_i^T\pa{\Omega_i(\gamma)}^{-1}\xi_i + \half\log\det\pa{\Omega_i(\gamma)}.
}

The derivative of the objective w.r.t $\gamma_j$, the $j$'th diagonal element of the matrix $\Gamma$ is
\eq{
	\pderiv{\xi_i^T\Omega_i^{-1}\xi_i}{\Gamma_{jj}} & = \trace \left[\left(\pderiv{\xi_i^T\Omega_i^{-1}\xi_i}{\Omega_i}\right)\pderiv{\Omega}{\Gamma_{jj}}\right] = \trace\left[\left(-\Omega_i^{-1}\xi_i\xi_i^T\Omega_i^{-1}\right)^T Z_i\pderiv{\Gamma}{\Gamma_{jj}}Z_i^T\right] 
}
where $\pderiv{\Gamma}{\Gamma_{jj}}$ is a single-entry matrix $J^{jj}$ with $jj$'th element is equal to 1 and zeroes elsewhere. 
Plugging in the value and making a circular swap, we get
Substituting this back we get
%\eq{
%	= \trace\left[\left(-\Omega_i^{-1}\xi_i\xi_i^T\Omega_i^{-1}\right)^T Z^j_i{Z^j_i}^T\right] = 
%} 
%where $Z^j_i$ is a $j$'th column of the matrix $Z_i$. Making a circular swap we end up with
\eq{
\pderiv{\xi_i^T\Omega_i^{-1}\xi_i}{\Gamma_{jj}} = \trace\left[-{Z^j_i}^T\Omega_i^{-1}\xi_i\xi_i^T\Omega_i^{-1} Z^j_i\right] = - ({Z_i^j}^T\Omega_i^{-1}\xi_i)^2
} 
Similarly,

\eq{
	\pderiv{\log\det\Omega_i}{\Gamma_{jj}} = \trace\left[\left(\pderiv{\log\det\Omega_i}{\Omega_i}\right)\pderiv{\Omega_i}{\Gamma_{jj}}\right] = \trace\left[\Omega_i^{-1}Z_i^j{Z_i^j}^T\right] = {Z_i^j}^T\Omega_i^{-1}Z_i^j
}
Using symmetry of $\Omega_i$, we have % is symmetric, we have 
%\eq{
%	\left[\nabla_\gamma \LL\pa{\beta, \gamma}\right]_j = \sum_{i = 1}^m  - ({Z_i^j}^T\Omega_i^{-1}\xi_i)^2 + {Z_i^j}^T\Omega_i^{-1}Z_i^j = 
%}
%or, in vector form
\eq{
\label{dgamma}
\nabla_\gamma \LL\pa{\beta, \gamma}
    = \sum_{i = 1}^m  \diag{({Z_i}^T\Omega_i^{-1}Z_i)} - ({Z_i}^T\Omega_i^{-1}\xi_i)^{\circ 2} 
}
where $\circ$ denotes the Hadamard (element-wise) product
and $\diag{\cdot}$ takes a square matrix to its diagonal. 
Using the Cholesky decomposition $\Omega_i = L_iL_i^T$ we can calculate~\eqref{dgamma} using only one triangular matrix inversion:
\eq{
\nabla_\gamma \LL\pa{\beta, \gamma}
    = \sum_{i = 1}^m \left[ \sum_{\text{rows}}\left(L_i^{-1}Z_i\right)^{\circ 2} -  [(L_i^{-1}Z_i)^{T}(L_i^{-1}\xi_i)]^{\circ 2}\right]
}
Notice, that the loss function (\ref{eq:lmm_objective}) and the optimal $\beta$ can also be effectively computed using Cholesky:

\eq{
	\LL\pa{\gamma} = \sum_{i = 1}^m \half\xi_i^T\pa{\Omega_i(\gamma)}^{-1}\xi_i + \half\log\det\pa{\Omega_i(\gamma)} = \sum_{i = 1}^m \half\ \|L_i^{-1}\xi_i\|^2 - \sum_{j = 1}^k \log{[L^{-1}_i]}_{jj}
}

\eq{
	\beta_{k+1} & = \argmin_{\beta}\LL(\beta, \gamma_{k}) = \left(\sum_{i = 1}^m X_i^T\Omega_i^{-1}X_i\right)^{-1}\sum_{i = 1}^mX_i^T\Omega_i^{-1}y_i = \\
	& = \left(\sum_{i = 1}^m (L_i^{-1}X_i)^TL_i^{-1}X_i\right)^{-1}\sum_{i = 1}^m (L_i^{-1}X_i)^TL_i^{-1}y_i 
}

The Hessian w.r.t. $\gamma$ is derived below:
\eq{
	\frac{\partial^2 \LL\pa{\beta, \gamma}}{\partial \gamma_j^2} & = \sum_{i=1}^m -2({Z_i^j}^T\Omega_i^{-1}\xi_i)\trace\left[\frac{\partial{Z_i^j}^T\Omega_i^{-1}\xi_i}{\partial \Omega_i}\frac{\partial \Omega_i}{\partial \Gamma_{jj}} \right] + \trace\left[\frac{\partial{Z_i^j}^T\Omega_i^{-1}Z_i^j}{\partial \Omega_i}\frac{\partial \Omega_i}{\partial \Gamma_{jj}} \right] = \\
	& = \sum_{i=1}^m 2({Z_i^j}^T\Omega_i^{-1}\xi_i)\trace\left[\Omega_i^{-1}Z_i^j{\xi_i}^T\Omega_i^{-1}Z_i^j{Z_i^j}^T \right] - ({Z_i^j}^T\Omega_i^{-1}Z_i^j)^2 = \\
	& = \sum_{i=1}^m 2({Z_i^j}^T\Omega_i^{-1}\xi_i)({Z_i^j}^T\Omega_i^{-1}Z_i^j)({\xi_i}^T\Omega_i^{-1}Z_i^j) - ({Z_i^j}^T\Omega_i^{-1}Z_i^j)^2
}

\eq{
	\frac{\partial^2 \LL\pa{\beta, \gamma}}{\partial \gamma_j\partial \gamma_k} & = \sum_{i=1}^m -2({Z_i^j}^T\Omega_i^{-1}\xi_i)\trace\left[\frac{\partial{Z_i^j}^T\Omega_i^{-1}\xi_i}{\partial \Omega_i}\frac{\partial \Omega_i}{\partial \Gamma_{kk}} \right] + \trace\left[\frac{\partial{Z_i^j}^T\Omega_i^{-1}Z_i^j}{\partial \Omega_i}\frac{\partial \Omega_i}{\partial \Gamma_{kk}} \right] = \\
	& = \sum_{i=1}^m 2({Z_i^j}^T\Omega_i^{-1}\xi_i)\trace\left[\Omega_i^{-1}Z_i^j{\xi_i}^T\Omega_i^{-1}Z_i^k{Z_i^k}^T \right] - ({Z_i^j}^T\Omega_i^{-1}Z_i^k)^2 = \\
	& = \sum_{i=1}^m 2(\xi_i^T\Omega_i^{-1}{Z_i^j})({Z_i^j}^T\Omega_i^{-1}Z_i^k)({Z_i^k}^T\Omega_i^{-1}{\xi_i}) - ({Z_i^j}^T\Omega_i^{-1}Z_i^k)^2
}

\eq{
	\label{eq:lmm_hessian_gamma}
	\nabla^2_\gamma\LL\pa{\beta, \gamma} = \half\sum_{i = 1}^m -(Z_i^T\Omega_i^{-1}Z_i)^{\circ 2}  + 2\Diag{Z_i^T\Omega_i^{-1}\xi_i}(Z_i^T\Omega^{-1}Z_i)\Diag{\xi_i^T\Omega^{-1}Z_i} = \\
	= \half\sum_{i = 1}^m -(Z_i^T\Omega_i^{-1}Z_i)^{\circ 2} + 2(Z_i^T\Omega_i^{-1}\xi_i)(\xi_i^T\Omega^{-1}Z_i)^T\circ(Z_i^T\Omega^{-1}Z_i)
}

\subsection{Derivation of Selected Proximal Operators from Table \ref{table:proxes}}
\label{ch:proxes_appendix}

\paragraph{SCAD}

For a scalar variable $x \in \R$, SCAD-regularizer is defined as 
\eq{
r(x) = \begin{cases} 
    \sigma |x|, & |x| \leq \sigma \\
    \frac{-x^2 + 2\rho\sigma |x| - \sigma^2}{2(\rho - 1)}, & \sigma < |x| < \rho\sigma \\ \frac{\sigma^2(\rho + 1)}{2}, & |x| > \rho\sigma 
    \end{cases}
}
To evaluate the $\prox_{\alpha r}$ operator we need to solve the following minimization problem:

\eq{
    \label{eq:prox_separated_minimization}
    \min_{x} r(x) + \frac{1}{2\alpha}(x - z)^2
}

For $\alpha = 1$, the solution was derived by \cite{fan1997comments}. 
Here we extend it  for an arbitrary $\alpha$. %to use it with PGD algorithm, this derivation closely follows its original source; we provide it for the sake of completeness. 
To identify the set of stationary points $\{x^*\}$ of a non-smooth function $f(x)$, we the optimality condition
\eq{
    0 \in \partial_x f(x^*)
}
where $\partial_x f(x)$ denotes a sub-differential set of $f$ at the point $x$. For the prox problem, we get % \ref{eq:prox_separated_minimization}  yields:
\eq{
    0 \in \frac{1}{\alpha}(x^* - z) + \partial r(x)_{x = x^*}
}
Since $r(x)$ is piece-wise defined the precise value of $\partial r(x)_{x = x^*}$ will depend on $x^*$:
\begin{enumerate}
    \item Let $0 < x^* \leq \sigma$, then we have $\partial r(x)_{x = x^*} = \{x^*\}$ and so
    \eq{
        x = z - \sigma\alpha, \quad z \in [\sigma\alpha, \sigma + \sigma\alpha]
    }
    \item Let $-\sigma\alpha \leq x^* < 0$, then we have $\partial r(x)_{x = x^*} = \{-x^*\}$ and so
    \eq{
        x = z + \sigma\alpha, \quad z \in [-\sigma - \sigma\alpha, -\sigma\alpha]
    }
    \item Let $x^* = 0$, then $\partial r(x)_{x = x^*} = [-1, 1]$, which yields
    \eq{
        \frac{1}{\alpha}(x^* - z) \in -\sigma[-1, 1] \quad \Rightarrow \quad      z \in [-\sigma\alpha, \sigma\alpha]
    }
%    or
%%    \eq{
%%        x^* \in z + [-\sigma\alpha, \sigma\alpha]
%%    }
%%    Given that $x^* = 0$, it provides us with the condition on $z$:
%    \eq{
%        z \in [-\sigma\alpha, \sigma\alpha]
%    }
    \item Let $\sigma < x^* < \rho\sigma$, then $r(x)_{x = x^*} = \frac{-{x^*}^2 + 2\rho\sigma x^* - \sigma^2}{2(\rho - 1)}$, which gives us 
    \eq{
        \frac{1}{\alpha}(x^* - z) = \frac{x^* - \rho\sigma}{\rho - 1}
    }
    To ensure that the stationary point is indeed a minimizer, we need to ensure that 
    \eq{ 
        \frac{1}{\alpha} - \frac{1}{\rho - 1} > 0 \quad \Rightarrow \quad \alpha < \rho - 1.
    }
Rearranging the terms to express $x^*$ as a function of $z$ we get % the value of prox minimizer:
    \eq{
        x^* = \frac{(\rho - 1)z - \lambda\rho\sigma}{\rho - 1 - \alpha} \quad \Rightarrow \quad         z \in [\sigma + \alpha\sigma, \rho\sigma]
    }
%    Flipping this expression back to $z$ being a function of $x^*$ we get
%    \eq{
%        z = \frac{x^*(\rho - 1 + \lambda) + \alpha\rho\sigma}{\rho - 1} \in\left[\frac{\sigma(\rho - 1 - \alpha) + \alpha\rho\sigma}{\rho - 1}, \frac{\sigma\rho(\rho - 1 - \alpha) + \alpha\rho\sigma}{\rho - 1} \right]
%    }
%    which, after algebraic simplifications, becomes
%    \eq{
%        z \in [\sigma + \alpha\sigma, \rho\sigma]
%    }
    \item Let $-\rho\sigma < x^* < -\sigma$, then, similarly to the previous case, we get 
    \eq{
        \frac{1}{\alpha}(x^* - z) = \frac{x^*+\rho\sigma}{\rho - 1}
    }
    %where the sign of the second term in the denominator changes due presence of absolute value. 
    Rearranging the terms to express $x$ in terms of $z$ we get:
    \eq{
        x^* = \frac{(\rho - 1)z + \lambda\rho\sigma}{\rho - 1 - \alpha} \quad \Rightarrow \quad       z \in [-\sigma -\alpha\sigma, -\sigma]
    }
%    Rearranging the terms to express $z$ in terms of $x^*$, and using that $-\sigma\rho < x^* < \sigma$ we get \eq{
%        z \in [-\sigma -\alpha\sigma, -\sigma]
%    }
    \item Finally, when $|x^*| \geq \sigma\rho$ we have $\partial r(x)_{x = x^*} = \{0\}$ and so 
    \eq{
        x^* = z, \quad |z| \geq \sigma\rho
    }
    Bundling all six cases together,  we have
    %when $\alpha + 1  < \rho$ and we get 
    \eq{
    \label{eq:scadProx}
        \prox_{\alpha r}(z) = \begin{cases} \sign(z)(|z| - \sigma\alpha)_+, & |z| \leq \sigma(1+\alpha) \\ \frac{(\rho - 1)z - \sign(z)\rho \sigma\alpha}{\rho - 1 - \alpha}, & \sigma(1+\alpha) < |z| \leq \max(\rho,1+\alpha)\sigma \\ z, & |z| > \max(\rho, 1+\alpha)\sigma \end{cases}
    }
The middle branch is active only when $\rho > 1+\alpha$. One special case of this is when $\alpha = 1$, and then~\eqref{eq:scadProx} recovers    the classic result by \cite{Fan2001}. 
    
To get $\prox_{\alpha r + \delta_{\R_+}}(z)$ from $\prox_{\alpha r}(z)$ we only need to notice that (1) the minimizer $x^*$ of 
\eq{
    \min_x r(x) + \delta_{\R_+} + \frac{1}{\alpha}(x - z)^2
}
can never be negative, and that (2) when the minimizer $x^*$ is exactly zero we get:
\eq{
    \frac{1}{\alpha}(x^* - z) \in -\partial(r(x)|_{x = x^*} + \delta_{\R_+}(x)|_{x = x^*})% = -[-\infty, 0] + [-\sigma, \sigma]) = [-\sigma, +\infty]
    \quad \Rightarrow \quad     z \in [-\infty, \sigma\alpha]
}
%which gives us the condition on $z$:
%\eq{
%    z \in [-\infty, \sigma\alpha]
%}
%Similarly, when $\C$ also includes the upper-bound constraint for REML likelihood, the ``largest'' branch $|x^*| \geq \rho$ gets affected. Namely, when $\rho\sigma \leq |x^*| < \bar\gamma$ we still get $\partial r(x)_{x = x^*} = \{0\}$ and so
%\eq{
%    x^* = z,  \sigma\rho \leq \quad |z| < \bar\gamma
%}
%However, in case of $x^* = \bar\gamma$ we get
%\eq{
%    \frac{1}{\alpha}(x^* - z) \in -\partial(r(x)|_{x = x^*} + \delta_{\R_{\leq \bar\gamma}}(x)|_{x = x^*}) = -(\{0\} + [0, +\infty])  = [-\infty, 0]
%}
%which happens when $z \geq \bar\gamma$. 
    
\end{enumerate}

\paragraph{A-LASSO}

A-LASSO regularizer is defined as 
\eq{
    r(x) = w|x|
}
where $w=1/|\hat{x}|$ with $\hat{x}$ the solution of a non-regularized problem (\cite{Zou2006}). 
%For the purposes of deriving the proximal operator of A-LASSO, $w$ is a constant and positive scalar value. 
The derivation of the proximal operator of A-LASSO nearly matches the steps 1, 2, and 3 that of SCAD above. We wish to evaluate
\eq{
    \min_{x} w|x| + \frac{1}{2\alpha}(x - z)^2
}
as a function of $z$.
The sub-differential optimality criterion yields
\eq{
    0 \in \frac{1}{\alpha}(x^* - z) + w\partial |x|
}
\begin{enumerate}
\item Let $0 < x^*$, then we have $\partial r(x)_{x = x^*} = \{x^*\}$ and so
    \eq{
        x^* = z - \alpha w, \quad z > \alpha w
    }
    \item Let $x^* < 0$, then we have $\partial r(x)_{x = x^*} = \{-x^*\}$ and so
    \eq{
        x^* = z + \alpha w, \quad z < -\alpha w
    }
    \item Let $x^* = 0$, then $\partial r(x)_{x = x^*} = [-1, 1]$, which yields
    \eq{
        \frac{1}{\alpha}(x^* - z) \in [-w, w] \quad \Rightarrow \quad      z \in [-\alpha w, \alpha w]
    }
%    or
%    \eq{
%        x^* \in z + [-\alpha w, \alpha w]
%    }
%    Given that $x^* = 0$, it yields the condition on $z$:
%    \eq{
%        z \in [-\alpha w, \alpha w]
%    }
\end{enumerate}
Combining all cases together we get
\eq{
    \prox_{\alpha r}(z) = \sign(z)(|z| - \alpha w)_+
}
Finally, $\prox_{\alpha r + \delta_{\R}}(z)$ can be derived by noticing that, in this case, (1) $x^* \geq 0$, and (2) when $x^* = 0$ the sub-differential changes due to the presence of the delta-function:
\eq{
    x^* = 0 \implies \frac{1}{\alpha}(x^* - z) \in -([-\alpha w, \alpha w] + [-\infty, 0]) = [-\alpha w, +\infty]
}
which gives us the condition 
\eq{
    x^* = z, \quad z \in [-\infty, \alpha w].
}

%Similar situation arises when $\C$ includes the upper-bound box constraint for REML. For $|x^*| = \bar\gamma$ we get that 
%\eq{
%    x^* = 0 \implies \frac{1}{\alpha}(x^* - z) \in -(w + [0, +\infty]) = [-\alpha w, +\infty]
%}
%which gives us the condition 
%\eq{
%    x^* = \bar\gamma, \text{ when } \quad z \geq \bar\gamma + \alpha w
%}

\paragraph{LASSO} LASSO is a particular case of A-LASSO above when $w = 1$. 

\paragraph{$\ell_0$-regularizer} Comparing to its counterparts above, the regularizer $R(x) = \delta_{\|x\|\leq k}(x)$ is non-separable. However, the proximal operator of it can still be evaluated analytically:
\eq{
    \left[\prox_{\alpha R}(z)\right]_i = \left[\argmin_{\|x\| \leq k} \frac{1}{2\alpha}\|x - z\|^2\right]_i = \begin{cases}z_i, & i \in \II_{k}\\ 0, & \text{ otherwise}\end{cases}
}
where $\II_k$ is a set of $k$ largest in their absolute value coordinates of $z$. To get $\prox_{\alpha R + \delta_{\R_+}}$ we replace $\II_k$ with a set of $k$ largest positive coordinates of $z$, and set the rest of the coordinates to $0$.

\subsection{Lipschitz-constant for Likelihood of a Linear Mixed-Effects Model}
\label{sec:lipschitz_constant}
Recall that a function $\LL(x)$ is called L-Lipschitz smooth when 

\eq{
    \|\nabla \LL(x) - \nabla \LL(y)\|_2 \leq L\|x-y\|_2
}

To find the Lipschitz-constant of the function $\LL_{ML}$ (\ref{eq:lmm_objective}) we will use the fact that $\LL(x)$ is L-Lipschitz if and only if $\|\nabla^2 \LL(x)\| \leq L$ for any $x$. Hence, to upper-bound L we need to upper-bound the norms of Hessians. % from (\ref{eq:all_derivatives}).
Assume that $\|y_i - X_i\beta\| \leq \eta$ where $\eta > 0$. We get % Using the same technique as in (\ref{eq:trick_with_norm}), we get that 

\eq{
    \left\|\nabla^2\LL(x)\right\|_2 & = \left\|\begin{bmatrix} \nabla^2_{\beta\beta}\LL(\beta, \gamma) & \nabla^2_{\beta\gamma}\LL(\beta, \gamma) \\ \nabla^2_{\gamma\beta}\LL(\beta, \gamma) & \nabla^2_{\gamma\gamma}\LL(\beta, \gamma) \end{bmatrix} \right\| \leq \sum_{i=1}^m \left\| \begin{bmatrix}\frac{\|X_i\|_2^2}{\|\Lambda_i\|_2} & \frac{\eta\|X_i\|_2\|Z_i\|^2_2}{\|\Lambda_i\|^2} \\ \frac{\eta\|X_i\|_2\|Z_i\|^2_2}{\|\Lambda_i\|^2} & \frac{\eta\|Z_i\|_2^4}{\|\Lambda_i\|_2^3} \end{bmatrix} \right\| \\  & \leq \sum_{i=1}^m\max\left(\frac{\|X_i\|_2^2}{\|\Lambda_i\|_2},  \frac{\eta\|X_i\|_2\|Z_i\|^2_2}{\|\Lambda_i\|^2},  \frac{\eta\|X_i\|_2\|Z_i\|^2_2}{\|\Lambda_i\|^2}, \frac{\eta\|Z_i\|_2^4}{\|\Lambda_i\|_2^3}\right) = L
}

\section{Description of Datasets and Experiments}
Table \ref{table:detailed_comparison_of_algorithms} provides a more detailed overview of the relative performance of the algorithms from the Table \ref{table:comparison_of_algorithms}.

\subsection{Detailed Results from Simulation from Table \ref{table:comparison_of_algorithms}}
\label{ch:appendix_detailed_comparison}

\begin{table}[h!]
    \centering
    \resizebox{\columnwidth}{!}{
    \begin{tabular}{llllll}
\toprule
    & Regularizer &                   L0 &                   L1 &               ALASSO &                 SCAD \\
Model & Metric &                      &                      &                      &                      \\
\midrule
PGD & Accuracy &           89 (75-95) &           73 (68-82) &           88 (72-98) &           71 (62-78) \\
    & FE Accuracy &           88 (70-95) &           56 (45-70) &          84 (65-100) &           53 (45-65) \\
    & RE Accuracy &          90 (75-100) &          91 (80-100) &          92 (80-100) &          89 (75-100) \\
    & F1 &           88 (74-95) &           77 (71-83) &           88 (74-97) &           75 (68-80) \\
    & FE F1 &           87 (72-95) &           67 (62-75) &          85 (70-100) &           66 (62-72) \\
    & RE F1 &          89 (74-100) &          91 (78-100) &          91 (78-100) &          88 (74-100) \\
    & Time &  25.73 (17.21-38.69) &  23.37 (16.35-32.26) &  20.24 (14.12-28.09) &  48.50 (30.66-79.67) \\
    & Iterations &  29662 (20985-43234) &  31693 (22361-45603) &  28912 (20915-39210) &  41724 (26911-69881) \\
MSR3 & Accuracy &           92 (75-98) &          88 (65-100) &          91 (75-100) &          92 (75-100) \\
    & FE Accuracy &          92 (70-100) &          84 (50-100) &          90 (70-100) &          93 (75-100) \\
    & RE Accuracy &          92 (80-100) &          91 (75-100) &          92 (75-100) &          90 (75-100) \\
    & F1 &           92 (76-97) &          88 (71-100) &          91 (74-100) &          91 (77-100) \\
    & FE F1 &          92 (75-100) &          87 (64-100) &          91 (75-100) &          94 (76-100) \\
    & RE F1 &          91 (78-100) &          90 (74-100) &          91 (74-100) &          89 (71-100) \\
    & Time &     0.08 (0.06-0.11) &     0.08 (0.06-0.11) &     0.07 (0.06-0.09) &     0.10 (0.07-0.13) \\
    & Iterations &           34 (28-43) &           35 (27-50) &           33 (27-39) &           44 (30-59) \\
\bottomrule
\end{tabular}
    }
    \caption{Comparison of performance of algorithms}
    \label{table:detailed_comparison_of_algorithms}
\end{table}

\subsection{GBD Bullying Data}
\label{appendix:bullying_covariates}
The author acknowledges his colleague and collaborator Damian Santomauro\footnote{\href{mailto:d.santomauro@uq.edu.au}{d.santomauro@uq.edu.au}, Affiliate Assistant Professor of Health Metrics Sciences, Institute for Health Metrics and Evaluation, University of Washington} for providing the dataset, the description of its covariates, and the expert assessment of their historical importance in different rounds of GBD study below.
\begin{enumerate}
	\item \texttt{cv\_symptoms}
	\begin{itemize}
		\item 0 = study assesses participants for MDD or anxiety disorders via a diagnostic interview to determine whether they have a diagnosis. 
		\item 1 = study uses a symptom scale (e.g., Beck Depression Inventory) and uses an established cut-off on that scale to determine caseness. 
		\item Has not  been significant in the past. 
	\end{itemize}
	\item \texttt{cv\_unadjusted}
	\begin{itemize}
		\item 0 = RR is adjusted for potential confounders (e.g., SES, etc.)	
		\item 1 = RR is not adjusted for potential confounders
		\item Has been significant in the past.
	\end{itemize}
	\item \texttt{cv\_b\_parent\_only} 
	\begin{itemize}
		\item 0 = Child is involved in reporting their own exposure to bullying.
		\item 1 = Only parent is involved in reporting the child’s exposure to bullying
		\item Has recently been significant. 
	\end{itemize}
	\item \texttt{cv\_or}
	\begin{itemize}
		\item 0 = estimate is a RR
		\item 1 = estimate is an odds ratio (OR)
		\item ORs are always larger than RRs; covariate may not be significant. 
	\end{itemize}
	\item \texttt{cv\_multi\_reg}
	\begin{itemize}
		\item 0 = RR is the ratio of the rate of the outcome in persons exposed vs all persons unexposed (including persons exposed to low-threshold bullying victimization)
		\item 1 =  RRs are estimated via a logistic regression where exposure represented by 3 categories: 1) No exposure, 2) Occasional exposure, 3) Frequent exposure. The RR for occasional exposure will exclude participants with frequent exposure, and the RR for frequent exposure will exclude participants with occasional exposure. 
		\item Is expected to be significant.
	\end{itemize}
	\item \texttt{cv\_low\_threshold\_bullying}
	\begin{itemize}
		\item 0 = uses a ‘frequent’ exposure  threshold for classing someone as exposed to bullying.
		\item 1 = uses an ‘occasional’ exposure threshold for classing someone as exposed to bullying.
		\item Has been significant in the past. 
	\end{itemize}
	\item \texttt{cv\_anx}
	\begin{itemize}
		\item 0 = estimate represents risk for MDD
		\item 1 = estimate represents risk for anxiety disorders
	\end{itemize}
	\item \texttt{cv\_selection\_bias}
	\begin{itemize}
		\item 0 = < 15\% attrition at followup
		\item 1 = $\geqslant$ 15\% attrition at followup
		\item Has been significant in the past.
	\end{itemize}
	\item \texttt{Percent\_female}
	\begin{itemize}
		\item Indicates \% of sample in estimate that are female. 
	\end{itemize}
	\item \texttt{cv\_child\_baseline}
	\begin{itemize}
	    \item Indicates whether mid-age of sample is is above or below 13.
		\item Has not been significant in the past.
	\end{itemize}
\end{enumerate}

\end{document}